\documentclass[a4paper,12pt, epsfig]{article}
\usepackage{epsfig}
\usepackage{float}
\restylefloat{table}
\usepackage{amssymb}
\usepackage{amsfonts}
\usepackage{amsmath}
\usepackage{graphicx}
\usepackage{tikz}
\usepackage{subfigure}
\newskip\humongous \humongous=0pt plus 1000pt minus 1000pt

\newif\ifdtup

\catcode`\@=11

\@addtoreset{equation}{section}
\def\theequation{\thesection.\arabic{equation}}

\def\@normalsize{\@setsize\normalsize{15pt}\xiipt\@xiipt
\abovedisplayskip 14pt plus3pt minus3pt%
\belowdisplayskip \abovedisplayskip
\abovedisplayshortskip \z@ plus3pt%
\belowdisplayshortskip 7pt plus3.5pt minus0pt}

\def\small{\@setsize\small{13.6pt}\xipt\@xipt
\abovedisplayskip 13pt plus3pt minus3pt%
\belowdisplayskip \abovedisplayskip
\abovedisplayshortskip \z@ plus3pt%
\belowdisplayshortskip 7pt plus3.5pt minus0pt
\def\@listi{\parsep 4.5pt plus 2pt minus 1pt
     \itemsep \parsep
     \topsep 9pt plus 3pt minus 3pt}}

\relax

\catcode`@=12

\catcode`\@=11

\def\section{\@startsection{section}{1}{\z@}{3.5ex plus 1ex minus
   .2ex}{2.3ex plus .2ex}{\large\bf}}

\def\thesection{\arabic{section}}
\def\thesubsection{\arabic{section}.\arabic{subsection}}

\def\appendix{\setcounter{section}{0}
 \def\thesection{Appendix \Alph{section}}
 \def\thesubsection{\Alph{section}.\arabic{subsection}}
 \def\theequation{\Alph{section}.\arabic{equation}}}

\def\SymBoxes#1#2#3#4{\newdimen\un@t \un@t#3%
\raisebox{#1}{\rule{#2\un@t}{#4}\hskip-#2\un@t
\@tempdimb\un@t \advance\@tempdimb by-#4\@tempcntb#2\relax%
\@whilenum{\@tempcntb>0}\do{
\rule{#4}{\un@t}\hskip\@tempdimb \advance\@tempcntb by\m@ne}%
\hskip-#2\un@t \rule[\un@t]{#2\un@t}{#4}%
\rule[\un@t]{#4}{#4}\hskip-#4
\rule{#4}{\un@t}}\hskip-#4}                

\begin{document}
\newcommand{\bal}{\begin{align}}
\newcommand{\eal}{\end{align}}
\newcommand{\tpsi}{\tilde \psi}

\newcommand{\beq}{\begin{equation}}
\newcommand{\eeq}{\end{equation}}
\newcommand{\bea}{\begin{eqnarray}}
\newcommand{\eea}{\end{eqnarray}}
\newcommand{\beas}{\begin{eqnarray*}}
\newcommand{\eeas}{\end{eqnarray*}}
\newcommand{\defi}{\stackrel{\rm def}{=}}
\newcommand{\non}{\nonumber}
\newcommand{\bquo}{\begin{quote}}
\newcommand{\enqu}{\end{quote}}
\renewcommand{\(}{\begin{equation}}
\renewcommand{\)}{\end{equation}}
\def \eqn#1#2{\begin{equation}#2\label{#1}\end{equation}}
\def\IZ{{\mathbb Z}}
\def\IR{{\mathbb R}}
\def\IC{{\mathbb C}}
\def\IQ{{\mathbb Q}}
\def\de{\partial}
\def\Tr{ \hbox{\rm Tr}}
\def\H{ \hbox{\rm H}}
\def\HE{ \hbox{$\rm H^{even}$}}
\def\HO{ \hbox{$\rm H^{odd}$}}
\def\K{ \hbox{\rm K}}
\def\Im{ \hbox{\rm Im}}
\def\Ker{ \hbox{\rm Ker}}
\def\const{\hbox {\rm const.}}
\def\o{\over}
\def\im{\hbox{\rm Im}}
\def\re{\hbox{\rm Re}}
\def\bra{\langle}\def\ket{\rangle}
\def\Arg{\hbox {\rm Arg}}
\def\Re{\hbox {\rm Re}}
\def\Im{\hbox {\rm Im}}
\def\exo{\hbox {\rm exp}}
\def\diag{\hbox{\rm diag}}
\def\longvert{{\rule[-2mm]{0.1mm}{7mm}}\,}
\def\a{\alpha}
\def\dag{{}^{\dagger}}
\def\tq{{\widetilde q}}
\def\p{{}^{\prime}}
\def\W{W}
\def\N{{\cal N}}
\def\hsp{,\hspace{.7cm}}

\def\br{\nonumber\\}
\def\IZ{{\mathbb Z}}
\def\IR{{\mathbb R}}
\def\IC{{\mathbb C}}
\def\IQ{{\mathbb Q}}
\def\IP{{\mathbb P}}
\def \eqn#1#2{\begin{equation}#2\label{#1}\end{equation}}
\def\tDelta{{\tilde \Delta}}
\def\tlambda{{\tilde \lambda}}

\newcommand{\C}{\ensuremath{\mathbb C}}
\newcommand{\Z}{\ensuremath{\mathbb Z}}
\newcommand{\R}{\ensuremath{\mathbb R}}
\newcommand{\rp}{\ensuremath{\mathbb {RP}}}
\newcommand{\cp}{\ensuremath{\mathbb {CP}}}
\newcommand{\vac}{\ensuremath{|0\rangle}}
\newcommand{\vact}{\ensuremath{|00\rangle}                    }
\newcommand{\oc}{\ensuremath{\overline{c}}}
\begin{titlepage}
\begin{flushright}
SISSA 37/2010/EP
\end{flushright}
\bigskip
\def\thefootnote{\fnsymbol{footnote}}

\begin{center}
{\Large
{\bf
The Many Phases of Holographic Superfluids \\
\vspace{0.1in}
}
}
\end{center}

\bigskip
\begin{center}
{\large Daniel AREAN$^1$\footnote{\texttt{arean@sissa.it}}}, {\large
Pallab BASU$^2$\footnote{\texttt{pallab@phas.ubc.ca}}}, and {\large
Chethan
KRISHNAN$^1$\footnote{\texttt{krishnan@sissa.it}}}\\
\end{center}

\renewcommand{\thefootnote}{\arabic{footnote}}

\begin{center}
{$^1$\em  {SISSA and INFN - Sezione di Trieste\\ Via Bonomea 265;
I-34136 Trieste, Italy\\
}}
{\vskip 0.3in}
{$^2$\em  {University of British Columbia, \\
Vancouver, Canada, V6T 1Z1 }}

\end{center}

\noindent
\begin{center} {\bf Abstract} \end{center}
We investigate holographic superfluids 
in $AdS_{d+1}$ with $d=3,4$ in the non-backreacted approximation for various masses of the scalar field. 
In $d=3$ the phase structure is universal for all the masses that we consider: the critical temperature decreases as the superfluid velocity increases, and as it is cranked high enough, the order of the phase transition changes from second to first. Surprisingly, in $d=4$ we find that the phase structure is more intricate. For sufficiently high mass, there is always a second order phase transition to the normal phase, no matter how high the superfluid velocity. For some parameters, as we lower the temperature, this transition happens before a first order transition to a new superconducting phase. Across this first order transition, the gap in the transverse conductivity jumps from almost zero to about half its maximum value. We also introduce a double scaling limit where we can study the phase transitions (semi-)analytically in the large velocity limit. The results corroborate and complement our numerical results. In $d=4$, this approach has the virtue of being fully analytically tractable. 

\begin{center}
\end{center}

\vspace{1.6 cm}
\vfill

\end{titlepage}
\hfill{}
\bigskip

\tableofcontents

\setcounter{footnote}{0}
\section{Introduction}

\noindent
High-$T_c$ superconductivity\footnote{Typically high-$T_c$ superconductors are Copper oxides of heavy transition elements, called the Cuprates.} is a challenging theoretical problem because it involves a strongly coupled non-linear phenomenon, while theoretical physics is mostly about solving (weak perturbations of) linear equations. One way we might actually be able to solve high-$T_c$ superconductivity is if we were able to translate the strongly coupled problem into a weakly coupled different problem, so that we can get to work on it with the usual hammer and tongs.

The AdS/CFT correspondence \cite{Maldacena, GKP, Witten} offers such a translation for many strongly coupled systems in terms of a weakly coupled theory that involves gravity in anti-de Sitter space. It has recently been found that some of the generic features of superconducting phase transitions and condensates can in fact be reproduced using an Einstein-Maxwell theory coupled to a charged scalar field in an AdS-black hole background \cite{Gubser:2008px,HHH1}. This gravitational system has instabilities towards the formation of scalar hair, and this has a dual interpretation in terms of a superconducting/superfluid \footnote{The bulk $U(1)$ gauge field is dual to a global $U(1)$ on the boundary. So these solutions are perhaps more accurately described as superfluids, not superconductors. But for many physical questions, the fact that the $U(1)$ is gauged is irrelevant because the effective ``dressed'' mass of the electrons is large. The flow of current in a real superconductor is often nothing but superfluid flow.} phase transition.

Superconductivity implies that the conductivity is infinite for DC currents, which in turn means that the system can sustain a steady DC current in the absence of any applied electric field. So any gravitational system that purports to be dual to a superconductor should have states which can be identified as the duals of such constant current states.  The AdS/CFT map claims that the boundary values of bulk-fields are sources for the dual fields on the boundary. A current $J_x$ along one of the spatial directions $x$ on the boundary is therefore dual to a gauge field $A_x$ turned on in that direction. So the holographic dual of a supercurrent would be an AdS black hole with vector hair (which should be thought of as the excited current carrying state) on top of the hairy scalar black hole ground state (the superconducting vacuum).

In \cite{Basu, Herzog} such a holographic supercurrent/superfluid solution was constructed\footnote{Other related works include the non-abelian case \cite{Basu:2008bh}, inclusion of gravity backreaction \cite{Tisza}, fixed supercurrent  case \cite{Daniel}, rotating cases \cite{rot}, supercurrent vortices \cite{vort} and inhomogeneous case \cite{Keranen:2009ss}.} in  $AdS_4$ when the charged scalar field was conformally coupled. The purpose of this paper is to generalize it to various scalar masses and also consider $AdS_5$ as well. The mass-squareds of the scalars we consider range from positive, down to those that saturate the Breitenlohner-Freedman bound.

In $d=3$ we find that the system exhibits the same phase structure in all cases: in every case that we have investigated, the critical temperature for the superconducting phase transition decreases as one increases the superfluid velocity \footnote{This is the leading order (constant) piece in the falloff of $A_x$ at the boundary. See Section 2.}, and in fact changes from a second order to a first order transition at a ``special point'' in the phase diagram. This was the behavior found in \cite{Basu, Herzog} for the conformally coupled case \footnote{See \cite{Ammon:2009xh} for an interesting phase diagram in a similar anisotropic system, where a backreacted non-abelian holographic superconductor was considered.}. We find the same general behavior in all the cases where we have done explicit direct numerical simulation. These cases all have $m^2 \le 4$ (we set the AdS scale to unity). We also find evidence based on a semi-analytic/semi-numeric approach that the same is true for masses as high as $m^2=20$ around which the numerics become unstable. Our results should be contrasted with the results of \cite{Tisza} where backreaction effects where taken into account in the conformally coupled case in $d=3$. It was found that when the charge of the scalar is small enough (which is the opposite limit to where one can ignore the backreaction effects), the phase transition stays second order even at high superfluid velocity. Another complementary study is \cite{Daniel} where conformally coupled case has been discussed in canonical ensemble, {\em i.e.} in a fixed current ensemble.

In $d=4$, our results are much more surprising. For low mass the situation is similar to $d=3$ case. However for sufficiently high mass, we find that there is always a second order phase transition to the normal phase, for arbitrarily high superfluid velocity. In some intermediate regimes, as we lower the temperature, this transition arises at a higher temperature than a subsequent first order transition to a new superconducting phase. We describe this structure (which we call the Cave of Winds for reasons to do with the shape of the condensate curve) in more detail in the main text. We have checked these claims by direct numerical simulation of the condensate (and the free energy) for some values of the masses. We have also developed an approach that works in a scaling limit at high velocities, which is fully analytical in $d=4$. The results using that approach are valid for phase transitions from the normal phase. They indicate a second order phase transition at very high velocities (for high enough mass), and are consistent with our direct numerical results where they overlap. We can analytically determine the critical mass at which this change in phase structure happens, and it turns out to be approximately $m^2 \approx -2.457$. In Fig. \ref{phaseplot}, we present schematic diagrams of the various phase transition scenarios we have found.  

The next section serves the dual purpose of introducing the problem and describing the setup. Section 3, which is the core section of this paper, presents out numerical results as well as presents the details of the phase structure. We also devote a part of the section to developing and applying the (semi-)analytical approach that can be used to study the phases in a high-velocity/low-temperature scaling limit. Section 4 discusses aspects of the conductivity plots. An appendix that recapitulates some of the relevant aspects of phase transitions is provided.
\ifx\du\undefined
  \newlength{\du}
\fi
\setlength{\du}{15\unitlength}
\begin{figure}\label{phaseplot}
\begin{center}
\subfigure[{For $d=3$ and $d=4$ with lower values of $m^2$. The undotted curve is the line of second order transition which ends in a special/critical point. The dotted line is for first order transition. }]{\resizebox{10cm}{!}{
\begin{tikzpicture}
\pgftransformxscale{1.000000}
\pgftransformyscale{-1.000000}
\definecolor{dialinecolor}{rgb}{0.000000, 0.000000, 0.000000}
\pgfsetstrokecolor{dialinecolor}
\definecolor{dialinecolor}{rgb}{1.000000, 1.000000, 1.000000}
\pgfsetfillcolor{dialinecolor}
\pgfsetlinewidth{0.100000\du}
\pgfsetdash{}{0pt}
\pgfsetdash{}{0pt}
\pgfsetbuttcap
{
\definecolor{dialinecolor}{rgb}{0.000000, 0.000000, 0.000000}
\pgfsetfillcolor{dialinecolor}
\pgfsetarrowsend{stealth}
\definecolor{dialinecolor}{rgb}{0.000000, 0.000000, 0.000000}
\pgfsetstrokecolor{dialinecolor}
\draw (20.000000\du,19.950000\du)--(20.050000\du,5.800000\du);
}
\pgfsetlinewidth{0.100000\du}
\pgfsetdash{}{0pt}
\pgfsetdash{}{0pt}
\pgfsetbuttcap
{
\definecolor{dialinecolor}{rgb}{0.000000, 0.000000, 0.000000}
\pgfsetfillcolor{dialinecolor}
\pgfsetarrowsend{stealth}
\definecolor{dialinecolor}{rgb}{0.000000, 0.000000, 0.000000}
\pgfsetstrokecolor{dialinecolor}
\draw (20.000000\du,19.900000\du)--(35.950000\du,19.900000\du);
}
\pgfsetlinewidth{0.100000\du}
\pgfsetdash{}{0pt}
\pgfsetdash{}{0pt}
\pgfsetbuttcap
{
\definecolor{dialinecolor}{rgb}{0.000000, 0.000000, 0.000000}
\pgfsetfillcolor{dialinecolor}
\definecolor{dialinecolor}{rgb}{0.000000, 0.000000, 0.000000}
\pgfsetstrokecolor{dialinecolor}
\pgfpathmoveto{\pgfpoint{34.350037\du}{19.850094\du}}
\pgfpatharc{339}{295}{16.133565\du and 16.133565\du}
\pgfusepath{stroke}
}
\definecolor{dialinecolor}{rgb}{0.000000, 0.000000, 0.000000}
\pgfsetstrokecolor{dialinecolor}
\pgfpathmoveto{\pgfpoint{34.350037\du}{19.850094\du}}
\pgfpatharc{339}{295}{16.133565\du and 16.133565\du}
\pgfusepath{stroke}
\pgfsetlinewidth{0.100000\du}
\pgfsetdash{}{0pt}
\pgfsetmiterjoin
\pgfsetbuttcap
\definecolor{dialinecolor}{rgb}{0.000000, 0.000000, 0.000000}
\pgfsetfillcolor{dialinecolor}
\pgfpathmoveto{\pgfpoint{26.050000\du}{11.050000\du}}
\pgfpathcurveto{\pgfpoint{26.102149\du}{10.936397\du}}{\pgfpoint{26.267900\du}{10.874943\du}}{\pgfpoint{26.381502\du}{10.927092\du}}
\pgfpathcurveto{\pgfpoint{26.495105\du}{10.979241\du}}{\pgfpoint{26.556559\du}{11.144992\du}}{\pgfpoint{26.504410\du}{11.258594\du}}
\pgfpathcurveto{\pgfpoint{26.452262\du}{11.372197\du}}{\pgfpoint{26.286511\du}{11.433651\du}}{\pgfpoint{26.172908\du}{11.381502\du}}
\pgfpathcurveto{\pgfpoint{26.059305\du}{11.329354\du}}{\pgfpoint{25.997851\du}{11.163603\du}}{\pgfpoint{26.050000\du}{11.050000\du}}
\pgfusepath{fill}
\definecolor{dialinecolor}{rgb}{0.000000, 0.000000, 0.000000}
\pgfsetstrokecolor{dialinecolor}
\pgfpathmoveto{\pgfpoint{26.050000\du}{11.050000\du}}
\pgfpathcurveto{\pgfpoint{26.102149\du}{10.936397\du}}{\pgfpoint{26.267900\du}{10.874943\du}}{\pgfpoint{26.381502\du}{10.927092\du}}
\pgfpathcurveto{\pgfpoint{26.495105\du}{10.979241\du}}{\pgfpoint{26.556559\du}{11.144992\du}}{\pgfpoint{26.504410\du}{11.258594\du}}
\pgfpathcurveto{\pgfpoint{26.452262\du}{11.372197\du}}{\pgfpoint{26.286511\du}{11.433651\du}}{\pgfpoint{26.172908\du}{11.381502\du}}
\pgfpathcurveto{\pgfpoint{26.059305\du}{11.329354\du}}{\pgfpoint{25.997851\du}{11.163603\du}}{\pgfpoint{26.050000\du}{11.050000\du}}
\pgfusepath{stroke}
\pgfsetlinewidth{0.100000\du}
\pgfsetdash{{\pgflinewidth}{0.200000\du}}{0cm}
\pgfsetdash{{\pgflinewidth}{0.200000\du}}{0cm}
\pgfsetbuttcap
{
\definecolor{dialinecolor}{rgb}{0.000000, 0.000000, 0.000000}
\pgfsetfillcolor{dialinecolor}
\definecolor{dialinecolor}{rgb}{0.000000, 0.000000, 0.000000}
\pgfsetstrokecolor{dialinecolor}
\draw (19.950000\du,8.450000\du)--(25.900000\du,11.050000\du);
}
\definecolor{dialinecolor}{rgb}{0.000000, 0.000000, 0.000000}
\pgfsetstrokecolor{dialinecolor}
\draw (20.866334\du,8.850415\du)--(25.900000\du,11.050000\du);
\pgfsetlinewidth{0.100000\du}
\pgfsetdash{}{0pt}
\definecolor{dialinecolor}{rgb}{0.000000, 0.000000, 0.000000}
\pgfsetstrokecolor{dialinecolor}
\draw (19.950000\du,8.450000\du)--(20.133267\du,8.530083\du);
\definecolor{dialinecolor}{rgb}{0.000000, 0.000000, 0.000000}
\pgfsetstrokecolor{dialinecolor}
\draw (20.255445\du,8.583472\du)--(20.438711\du,8.663555\du);
\definecolor{dialinecolor}{rgb}{0.000000, 0.000000, 0.000000}
\pgfsetstrokecolor{dialinecolor}
\draw (20.560889\du,8.716943\du)--(20.744156\du,8.797026\du);
\definecolor{dialinecolor}{rgb}{0.000000, 0.000000, 0.000000}
\pgfsetstrokecolor{dialinecolor}
\node[anchor=west] at (24.400000\du,15.600000\du){Phase I};
\definecolor{dialinecolor}{rgb}{0.000000, 0.000000, 0.000000}
\pgfsetstrokecolor{dialinecolor}
\node[anchor=west] at (31.650000\du,12.200000\du){Normal};
\definecolor{dialinecolor}{rgb}{0.000000, 0.000000, 0.000000}
\pgfsetstrokecolor{dialinecolor}
\node[anchor=west] at (33.500000\du,17.650000\du){2nd order};
\definecolor{dialinecolor}{rgb}{0.000000, 0.000000, 0.000000}
\pgfsetstrokecolor{dialinecolor}
\node[anchor=west] at (24.700000\du,11.850000\du){Critical point};
\definecolor{dialinecolor}{rgb}{0.000000, 0.000000, 0.000000}
\pgfsetstrokecolor{dialinecolor}
\node[anchor=west] at (22.000000\du,8.250000\du){1st Order};
\definecolor{dialinecolor}{rgb}{0.000000, 0.000000, 0.000000}
\pgfsetstrokecolor{dialinecolor}
\node[anchor=west] at (35.250000\du,21.350000\du){};
\definecolor{dialinecolor}{rgb}{0.000000, 0.000000, 0.000000}
\pgfsetstrokecolor{dialinecolor}
\node[anchor=west] at (34.350000\du,21.000000\du){Temperature};
\definecolor{dialinecolor}{rgb}{0.000000, 0.000000, 0.000000}
\pgfsetstrokecolor{dialinecolor}
\node[anchor=west] at (17.400000\du,5.400000\du){Superfluid velocity};
\end{tikzpicture}
}} \\
\end{center}
\subfigure[{Phase diagram with $d=4$ and intermediate values of $m^2$. At high enough superfluid velocity, there is a first order transition between superconducting phase III and superconducting phase I (dotted line). There is always a second order  transition between normal phase and phase I (undotted line). The behaviour shown near $T=0$ is tentative.}]{\resizebox{9cm}{!}{
\begin{tikzpicture}
\pgftransformxscale{1.000000}
\pgftransformyscale{-1.000000}
\definecolor{dialinecolor}{rgb}{0.000000, 0.000000, 0.000000}
\pgfsetstrokecolor{dialinecolor}
\definecolor{dialinecolor}{rgb}{1.000000, 1.000000, 1.000000}
\pgfsetfillcolor{dialinecolor}
\pgfsetlinewidth{0.100000\du}
\pgfsetdash{}{0pt}
\pgfsetdash{}{0pt}
\pgfsetbuttcap
{
\definecolor{dialinecolor}{rgb}{0.000000, 0.000000, 0.000000}
\pgfsetfillcolor{dialinecolor}
\pgfsetarrowsend{stealth}
\definecolor{dialinecolor}{rgb}{0.000000, 0.000000, 0.000000}
\pgfsetstrokecolor{dialinecolor}
\draw (20.000000\du,19.950000\du)--(20.050000\du,5.800000\du);
}
\pgfsetlinewidth{0.100000\du}
\pgfsetdash{}{0pt}
\pgfsetdash{}{0pt}
\pgfsetbuttcap
{
\definecolor{dialinecolor}{rgb}{0.000000, 0.000000, 0.000000}
\pgfsetfillcolor{dialinecolor}
\pgfsetarrowsend{stealth}
\definecolor{dialinecolor}{rgb}{0.000000, 0.000000, 0.000000}
\pgfsetstrokecolor{dialinecolor}
\draw (20.000000\du,19.900000\du)--(35.950000\du,19.900000\du);
}
\pgfsetlinewidth{0.100000\du}
\pgfsetdash{}{0pt}
\pgfsetdash{}{0pt}
\pgfsetbuttcap
{
\definecolor{dialinecolor}{rgb}{0.000000, 0.000000, 0.000000}
\pgfsetfillcolor{dialinecolor}
\definecolor{dialinecolor}{rgb}{0.000000, 0.000000, 0.000000}
\pgfsetstrokecolor{dialinecolor}
\pgfpathmoveto{\pgfpoint{34.050215\du}{19.800306\du}}
\pgfpatharc{325}{296}{36.454032\du and 36.454032\du}
\pgfusepath{stroke}
}
\definecolor{dialinecolor}{rgb}{0.000000, 0.000000, 0.000000}
\pgfsetstrokecolor{dialinecolor}
\node[anchor=west] at (31.650000\du,12.200000\du){Normal};
\definecolor{dialinecolor}{rgb}{0.000000, 0.000000, 0.000000}
\pgfsetstrokecolor{dialinecolor}
\node[anchor=west] at (32.200000\du,16.900000\du){2nd order};
\pgfsetlinewidth{0.100000\du}
\pgfsetdash{{1.000000\du}{0.200000\du}{0.200000\du}{0.200000\du}{0.200000\du}{0.200000\du}}{0cm}
\pgfsetdash{{1.000000\du}{0.200000\du}{0.200000\du}{0.200000\du}{0.200000\du}{0.200000\du}}{0cm}
\pgfsetbuttcap
{
\definecolor{dialinecolor}{rgb}{0.000000, 0.000000, 0.000000}
\pgfsetfillcolor{dialinecolor}
\definecolor{dialinecolor}{rgb}{0.000000, 0.000000, 0.000000}
\pgfsetstrokecolor{dialinecolor}
\pgfpathmoveto{\pgfpoint{20.049641\du}{17.050096\du}}
\pgfpatharc{76}{-70}{4.455278\du and 4.455278\du}
\pgfusepath{stroke}
}
\definecolor{dialinecolor}{rgb}{0.000000, 0.000000, 0.000000}
\pgfsetstrokecolor{dialinecolor}
\node[anchor=west] at (24.350000\du,16.050000\du){Phase III};
\definecolor{dialinecolor}{rgb}{0.000000, 0.000000, 0.000000}
\pgfsetstrokecolor{dialinecolor}
\node[anchor=west] at (25.250000\du,15.900000\du){};
\definecolor{dialinecolor}{rgb}{0.000000, 0.000000, 0.000000}
\pgfsetstrokecolor{dialinecolor}
\node[anchor=west] at (20.300000\du,11.500000\du){Phase I};
\definecolor{dialinecolor}{rgb}{0.000000, 0.000000, 0.000000}
\pgfsetstrokecolor{dialinecolor}
\node[anchor=west] at (22.000000\du,8.400000\du){};
\definecolor{dialinecolor}{rgb}{0.000000, 0.000000, 0.000000}
\pgfsetstrokecolor{dialinecolor}
\node[anchor=west] at (34.500000\du,20.450000\du){};
\definecolor{dialinecolor}{rgb}{0.000000, 0.000000, 0.000000}
\pgfsetstrokecolor{dialinecolor}
\node[anchor=west] at (33.200000\du,20.850000\du){Temperature};
\definecolor{dialinecolor}{rgb}{0.000000, 0.000000, 0.000000}
\pgfsetstrokecolor{dialinecolor}
\node[anchor=west] at (19.300000\du,5.500000\du){Superfluid velocity};
\definecolor{dialinecolor}{rgb}{0.000000, 0.000000, 0.000000}
\pgfsetstrokecolor{dialinecolor}
\node[anchor=west] at (20.700000\du,9.450000\du){?};
\end{tikzpicture}
}}\hspace{0.2in}
\subfigure[{Phase diagram for $d=4$ with large values of $m^2$. Only a second order transition is possible between superconducting and normal phase.}]{\resizebox{9cm}{!}{\begin{tikzpicture}
\pgftransformxscale{1.000000}
\pgftransformyscale{-1.000000}
\definecolor{dialinecolor}{rgb}{0.000000, 0.000000, 0.000000}
\pgfsetstrokecolor{dialinecolor}
\definecolor{dialinecolor}{rgb}{1.000000, 1.000000, 1.000000}
\pgfsetfillcolor{dialinecolor}
\pgfsetlinewidth{0.100000\du}
\pgfsetdash{}{0pt}
\pgfsetdash{}{0pt}
\pgfsetbuttcap
{
\definecolor{dialinecolor}{rgb}{0.000000, 0.000000, 0.000000}
\pgfsetfillcolor{dialinecolor}
\pgfsetarrowsend{stealth}
\definecolor{dialinecolor}{rgb}{0.000000, 0.000000, 0.000000}
\pgfsetstrokecolor{dialinecolor}
\draw (20.000000\du,19.950000\du)--(20.050000\du,5.800000\du);
}
\pgfsetlinewidth{0.100000\du}
\pgfsetdash{}{0pt}
\pgfsetdash{}{0pt}
\pgfsetbuttcap
{
\definecolor{dialinecolor}{rgb}{0.000000, 0.000000, 0.000000}
\pgfsetfillcolor{dialinecolor}
\pgfsetarrowsend{stealth}
\definecolor{dialinecolor}{rgb}{0.000000, 0.000000, 0.000000}
\pgfsetstrokecolor{dialinecolor}
\draw (20.000000\du,19.900000\du)--(35.950000\du,19.900000\du);
}
\pgfsetlinewidth{0.100000\du}
\pgfsetdash{}{0pt}
\pgfsetdash{}{0pt}
\pgfsetbuttcap
{
\definecolor{dialinecolor}{rgb}{0.000000, 0.000000, 0.000000}
\pgfsetfillcolor{dialinecolor}
\definecolor{dialinecolor}{rgb}{0.000000, 0.000000, 0.000000}
\pgfsetstrokecolor{dialinecolor}
\pgfpathmoveto{\pgfpoint{34.050215\du}{19.800306\du}}
\pgfpatharc{325}{296}{36.454032\du and 36.454032\du}
\pgfusepath{stroke}
}
\definecolor{dialinecolor}{rgb}{0.000000, 0.000000, 0.000000}
\pgfsetstrokecolor{dialinecolor}
\node[anchor=west] at (23.950000\du,14.600000\du){Phase I};
\definecolor{dialinecolor}{rgb}{0.000000, 0.000000, 0.000000}
\pgfsetstrokecolor{dialinecolor}
\node[anchor=west] at (28.200000\du,11.300000\du){Normal phase};
\definecolor{dialinecolor}{rgb}{0.000000, 0.000000, 0.000000}
\pgfsetstrokecolor{dialinecolor}
\node[anchor=west] at (32.700000\du,17.550000\du){2nd order};
\definecolor{dialinecolor}{rgb}{0.000000, 0.000000, 0.000000}
\pgfsetstrokecolor{dialinecolor}
\node[anchor=west] at (34.450000\du,21.100000\du){Temperature};
\definecolor{dialinecolor}{rgb}{0.000000, 0.000000, 0.000000}
\pgfsetstrokecolor{dialinecolor}
\node[anchor=west] at (17.950000\du,5.600000\du){Superfluid velocity};
\end{tikzpicture}
}}
\caption{Phase structure}
\end{figure}
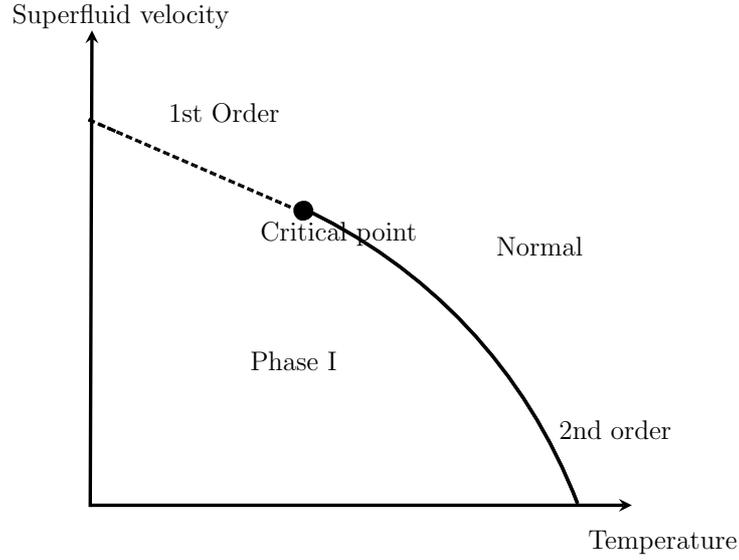
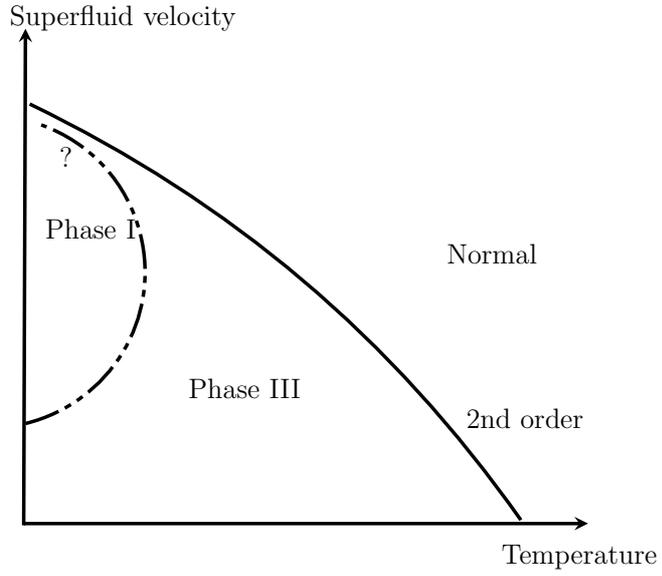
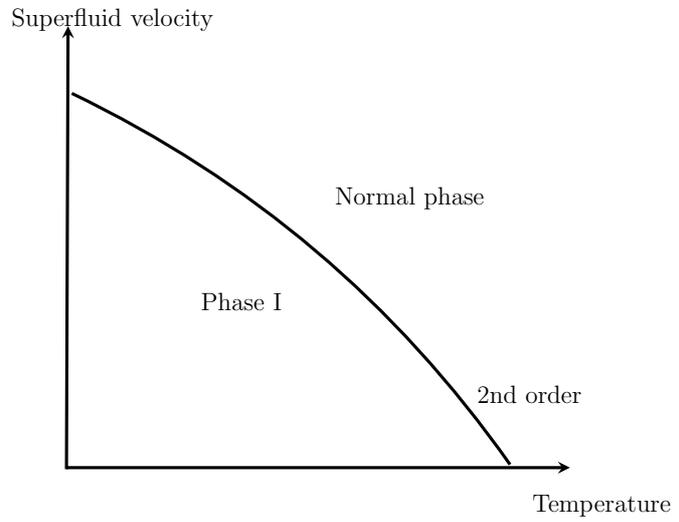
\section{Setup}

We will work with the black brane in AdS (dimension arbitrary for now) coupled to a Maxwell field and a charged scalar. In general dimensions, the matter piece in the action takes the form
\bea
S=\int d^{d+1}x\sqrt{-g} \left(-\frac{1}{4}\tilde F^{ab}\tilde F_{ab}-V(|\tilde \psi|)-|\nabla \tilde \psi-i q  \tilde A \tilde \psi|^2\right)\,.
\eea
We have set the gauge coupling $g=1$, which is merely a choice of unit. But the charge $q$ of the scalar is still arbitrary.  We would like to take a limit where we can treat the matter part of the action merely as a probe, so that the matter piece is negligible compared to the Einstein piece in the full action. To do this, we redefine $\Psi= q \tilde \psi$ and ${\cal A}= q \tilde A$, so that the matter action takes the form
\bea
S=\frac{1}{q^2}\int d^{d+1}x\sqrt{-g} \left(-\frac{1}{4} F^{ab} F_{ab}-V(| \Psi|)-|\nabla  \Psi-i   {\cal A}  \Psi|^2\right). \label{qaction}
\eea
Now it is clear that sending $q\rightarrow \infty$ while holding $\Psi, {\cal A}$ fixed takes us to the probe limit. In particular, the matter action dies relative to the Einstein action, so the full equations of motion for the metric will be the vacuum equations. We can then solve for  $\Psi, {\cal A}$ in the given Einstein vacuum solution, because the matter EOMs are unaffected by the overall factor of $1/q^2$ in the action above. By writing the complex scalar as $\Psi=\psi\, e^{i\theta}$ and doing the gauge transformation $A_b={\cal A}_b-\nabla_b \theta$, we can get rid of the phase $\theta$ and bring the equations of motion to the form
\bea
\nabla^aF_{ab}=2  \psi^2 A_b, \hspace{1.5in}\\
\nabla^a\nabla_a \psi -i (\nabla^aA_a) \psi-2i A^a\nabla_a\psi-(A^aA_a)\psi-\frac{1}{2}V'(\psi)=0.
\eea

The metric background we will work with is the planar black hole (AdS black brane):
\bea
ds^2=-f dt^2+f^{-1}dr^2+r^2 dx_idx_i, \ \ \ f(r)=r^2\Big(1-\frac{r_0^d}{r^d}\Big). \label{BHmetric}
\eea
We make the choice that the AdS radius is set to unity. The boundary of AdS is at $r \rightarrow \infty$ in these coordinates.  To turn on a supercurrent, following \cite{Basu,Herzog} we look for a bulk vector potential of the form $A_a=(A_t(r),0,A_x(r),0, ...)$, where the dots indicate that all of those components are zero. The equations of motion in the above background take the form
\bea
A_t''+\frac{d-1}{r}A_t'-\frac{2\psi^2}{f}A_t=0,  \hspace{0.75in} \\
\nonumber A_x''+\Big(\frac{f'}{f}+\frac{d-3}{r}\Big)A_x'-\frac{2\psi^2}{f}A_x=0, \hspace{0.4in}\\
\nonumber \psi''+\Big(\frac{f'}{f}+\frac{d-1}{r}\Big)\psi'+\Big(\frac{A_t^2}{f^2}-\frac{A_x^2}{r^2 f}\Big)\psi-\frac{1}{2}{V'(\psi) \over f}=0. \label{seom}
\eea
We will set $V(\psi)=m^2 \psi^2$ from now on, and try various values of the mass $m^2$ in what follows\footnote{In $d=3$ the choice $m^2=-2$ corresponds to the conformally coupled case considered in \cite{Basu, Herzog}.}. With this choice, the potential term in the EOMs becomes linear. 

For a generic mass $m^2$  the falloff of the scalar and the gauge field at the boundary can be determined by solving the asymptotic form of the equations above, which are nothing but the corresponding equations in pure AdS. One then gets the following falloffs:
\bea
\psi=\frac{\psi_-}{r^{\lambda_-}}+\frac{\psi_+}{r^{\lambda_+}}+ ..., \ \ \ \ {\rm where} \ \ \ \lambda_{\pm}=\frac{1}{2}(d\pm\sqrt{d^2+4m^2}), \label{falloff1}\\
A_t=\mu-\frac{\rho}{r^{d-2}}+..., \ \ \ \ A_x =S_x-\frac{J_x}{r^{d-2}}+...\,. \hspace{1in} \label{falloff2}
\eea
For the scalar, the usual AdS/CFT correspondence \cite{Witten} identifies the boundary value of the normalizable mode to a vev in the dual field theory and that of the non-normalizable mode to a source that couples to the vev (we discuss some subtleties in this statement momentarily). Our aim is to set the source term to zero and compute the vev by starting from the horizon and integrating out to the boundary. For $m^2 \geq -d^2/4+1$, only the $\lambda_+$ mode is normalizable, so there is no ambiguity: we set $\psi_-=0$ and identify $\psi_+ =\langle O\rangle$.  In the regime $-d^2/4 \leq m^2 \leq -d^2/4+1$ both modes are normalizable, so we can choose to set either to zero, and treat the other as the condensate. In the plots, we report both. For the boundary case $m^2=-d^2/4 $ where the scalar is at the Breitenlohner-Freedman bound, the two $\lambda$'s coincide and therefore there is a logarithmic branch. We set this logarithmic branch to zero because otherwise it triggers an instability \cite{Hubeny}.

\begin{table}[H]
 \begin{minipage}{.5\textwidth}
\centering
\begin{tabular}{|c|c|c|}\hline
            $m^2$ & $\lambda_+$ & $\lambda_-$
\\ [1ex] \hline
            $-{9\over 4}$&${3 \over 2}$&${3 \over 2}$\\[.5ex] \hline
                $-2$&$2 $&$1$\\[.5ex] \hline
                $0$&$3$&$0$\\[.5ex] \hline
        \end{tabular}
 \end{minipage}
 \begin{minipage}{.5\textwidth}
\centering
\begin{tabular}{|c|c|c|}\hline
            $m^2$ & $\lambda_+$ & $\lambda_-$
\\ [1ex] \hline
	$-4$&$2 $&$2$\\[.5ex] \hline
                $-{7 \over 4}$&${7 \over 2} $&${1 \over 2}$\\[.5ex] \hline
                $-{15\over 4}$&${5 \over 2}$&${3 \over 2}$\\[.5ex] \hline
                $-3$&$3$&$1$\\[.5ex] \hline
	$0$&$4$&$0$\\[.5ex] \hline
        \end{tabular}
 \end{minipage}
 \caption{The cases we consider in $d=3$ (left) and $d=4$ (right). We list the dimensions of both the normalizable and non-normalizable modes, but set the non-normalizable mode to zero to do the numerics. When both modes are normalizable, we consider both possibilities separately.}
    \label{tab2}
\end{table}

In each dimension, the minimal choices would be to consider (1) the case corresponding to the BL bound, (2) a case in the range where both modes are normalizable, and finally (3) a case where only one mode is normalizable. The cases we consider in gory detail are collected in Table 1. Besides these, we have also looked at some special cases less exhaustively. We also introduce a semi-analytic approach in a high velocity scaling limit, and consider a range of masses in both $d=3$ and $d=4$. This has the advantage of being simpler, but at the expense of loosing some information regarding the detailed structure. In $d=4$ near $m^2\approx -2.5$ the phases have a lot of structure and we have not attempted an exhaustive scan. In both $d=3$ and $d=4$, we consider the conformally coupled scalar\footnote{A generic scalar Lagrangian in a curved background is not Weyl invariant. But if the Lagrangian consists of the standard kinetic term and a mass-like term with ``mass" of the form  $\xi_{d+1} R$ where $R$ is the Ricci scalar, the action happens to be Weyl invariant if $\xi_{d+1}={(d-1) \over 4d}$. This is called a conformally coupled scalar. When the background is $AdS_{d+1}$, $R=-{d(d+1)\over L^2}$  (we have retained the AdS scale), and the ``mass"  $m^2=-{(d^2-1)\over 4L^2}$ is a constant and becomes a true mass. When we set $L=1$ this gives $m^2=-2$  in $d=3$ and $m^2=-{15 \over4}$ in $d=4$. } in detail. In $d=3$, this is the case with $m^2=-2$ and was already investigated in \cite{Basu, Herzog}.

The problem as we have stated so far has a scaling symmetry that we can use to simplify the numerics. We note that we can rescale
\bea
r \rightarrow r/r_0\;,\quad A_t \rightarrow A_t/r_0\;,\quad A_x \rightarrow A_x/r_0\;,\quad \psi \rightarrow \psi\;,\quad x_i \rightarrow x_i/r_0\;,
\label{r0scaling}
\eea
to get rid of the horizon location $r_0$ from the equations. In other words, we set the horizon location
at $r=1$. We also rescale the boundary coordinates to keep the boundary metric in the usual Minkowski form.

The gauge field component $A_t$ has to be zero in order for the norm of $A$ to be finite at the horizon \cite{HHH1}. Expanding the fields in a Taylor expansion around the horizon $r=1$ and plugging into the equations of motion using $A_t(r=1)=0$ we find  the constraints
\bea \label{constraint}
\psi'&=&\frac{1}{d}\,(A_x^2-m^2)\psi , \\
A_x'&=&\frac{2}{d}\, A_x\, \psi^2 ,
\eea
at the horizon.

We are interested in the dimensions of the various quantities in terms of the temperature of the black hole. The temperature $T$ goes linearly with $r_0$, so by using (\ref{r0scaling}) we find from  (\ref{falloff1}, \ref{falloff2}) the dimensions of the various quantities as follows:
\bea
[\psi_-]=\lambda_-\;,\quad [\psi_+]=\lambda_+\;,\quad [\mu]=[S_x]=1\;,\quad [\rho]=[J_x]=d-1\,.
\eea
Since $\mu$ has the dimensions of temperature, $T/\mu$ is a dimensionless quantity. As we have chosen to rescale $r_0$ to unity, we will work with $1/\mu$ instead of temperature. Before we proceed, a word about terminology: we call $S_x$ the superfluid velocity and $J_x$ the current. This is reasonable, because in conventional condensed matter physics, the gradient of the phase of the condensate is what is usually called the superfluid velocity \cite{Tinkham}. In turn, the gradient of the phase can be traded for the vector potential by a gauge transformation as discussed previously. So it is natural to associate the leading piece in the falloff of the vector potential to the superfluid velocity. Furthermore, the superfluid velocity can be thought of as the source of the supercurrent as dictated by AdS/CFT \cite{Witten}: the Lagrangian for the superfluid \cite{Tinkham} indeed has such a form.

The simple strategy in what follows is to integrate the EOMs from horizon to boundary. These are three second order equations, so we need six numbers to fix them. The horizon value of $A_t$ being zero and the resulting constraints (\ref{constraint}) fix three of them. Setting the source term in the scalar fixes another. So we are left with a two-parameter family of solutions. This means that there are secret relations between $\langle O\rangle \equiv \psi_+, T\equiv 1/\mu, S_x$ and $J_x$. What we are interested in is the plot of $\langle O\rangle$ in the $S_x, T$ plane and see how robust the phase structure is, as we tune the mass of the scalar and the dimension of spacetime.

\section{Phase Structure}

In this section, we present the results of the computations in $d=3, 4$ for various values of the mass and for various values of the superfluid velocity. As explained in the previous section, we will use $\mu$ to set the scale in the system. This means that we are working in the grand canonical ensemble, where the charge density (as measured in $\mu$) can vary, while the chemical potential (as measured in $\mu$) is fixed. We can work also in the canonical ensemble, where $\rho$ is what sets the scale of the system in which case we will be in the canonical ensemble. The general nature of the plots that we present are qualitatively unaffected by this switch (even though we have checked this explicitly only for a handful of cases). Each curve is for fixed $\tilde S\equiv S_x/\mu$, while scanning temperature, $T=1/\mu$.  When presenting the final result, we have found it useful to normalize the temperatures in terms of the critical temperature at zero velocity, $T_c$. The representative values of $\tilde S$ that we have chosen are noted on each plot. It is also important to note that the condensate axes are also appropriately scaled (as indicated on the plots) for convenience. To be precise, for an operator $O_\lambda$ of dimension $\lambda$ that condenses, we plot $\langle O_\lambda \rangle 
/T_c^{\lambda}$ vs. $T/T_c$.

Following \cite{Basu, Herzog} we study our system in a grand canonical ensemble, {\em i.e.} we keep
the superfluid velocity fixed when comparing between different solutions. In principle one may work in the canonical ensemble and keep
the supercurrent fixed \cite{Daniel}. In the paper by Horowitz and Roberts \cite{Horowitz}, various condensates of holographic superconductors were investigated. We have checked that our results precisely agree with their results when the superfluid velocity is zero.

\subsection{Instability and the Normal Phase}

The hairless solution to the equations (\ref{seom}) is given by
\begin{equation}
A_x=S_x, \ \  A_t=\mu \Big(1-\frac{1}{r^{d-2}}\Big), \ \  \psi=0.
\label{eqn:nbkg}
\end{equation}
This is the normal phase solution that exists for all temperatures. This solution is just the non-backreacted version of the RN solution ({\em i.e.} the backreaction of the gauge fields on the geometry is ignored). Our goal here is to argue that as we increase $\mu$, and thus decrease $T\equiv 1/\mu$, the scalar field in the above background develops a tachyonic mode and condenses. The result can be thought of as a new bound state in an auxiliary Schr\"odinger problem as we now explain.

The scalar EOM in (\ref{seom}) may be written as (assuming $V(\psi)=m^2 \psi^2$)
\beq
\psi''+\Big(\frac{f'}{f}+\frac{d-1}{r}\Big)\psi'+\Big(\frac{A_t^2}{f^2}-\frac{A_x^2}{r^2 f}\Big)\psi-{m^2 \over f}\psi=0\,.
\label{psieom}
\eeq
After applying the following change of variables:
\beq
\psi={\tilde\psi\over r^{d-1\over2}}\;,\qquad {dr\over dy}={1\over f}\,,
\label{psicov}
\eeq
the scalar EoM (\ref{psieom}) takes the form of a Schr\"odinger equation:
\beq
\frac{d^2}{dy^2} \tilde \psi-\tilde V_{\rm eff}(y)\tilde \psi = 0\,.
\label{psischreom}
\eeq
Notice that $y\rightarrow\infty$ as $r\rightarrow 1$ and $y\rightarrow 0$ as $r\rightarrow \infty$.
The potential $V_{\rm eff}$, written in terms of $r$, reads
\beq
V_{\rm eff}(r)=-f^2 \Big(-\frac{(d-1)(d-3)}{4r^2}-\frac{(d-1)f'}{2rf}+\frac{A_t^2}{f^2}-\frac{A_x^2}{r^2 f}-{m^2 \over f}\Big)\,.
\label{schrpot}
\eeq
The EOM of $\psi$ has been rephrased as a Schr\"odinger-type potential problem on a semi-infinite line ($y:[0,\infty]$). Depending on the nature of the potential $V_{\rm eff}$,  there may exist a bound state for $\psi$. A bound state such as this signifies an instability, suggesting that there may be new phases with non-trivial $\psi$. The resulting new phase (with non-trivial scalar and gauge field) is called a ``superconducting'' phase (and the solution is called a holographic superconductor) as it shows infinite DC conductivity \cite{HHH1}.

In the absence of $A_x$ and $A_t$, the scalar $\psi$ may condense if and only if $m^2 \leq m_{BF}^2$. This is the well known Breitenlohner-Freedman (BF) bound. Generically, by choosing a sufficiently negative $V_{\rm eff}$ one may force $\psi$ to condense, e.g. if we set $S_x=0,A_t=\mu(1-{1 \over r})$ and increase $\mu$ from zero then eventually 
the solution will become unstable. For even larger values of $\mu$, there will be a bound state of $\psi$. The existence of such a bound state may be rigorously shown using a trial wave function \cite{Hartnoll:2008kx}. The situation is a little more complicated if $S_x \neq 0$. It is clear from the nature of the potential that a non-zero $S_x$ discourages condensation. Using the method of trial wave function it may be possible to address this problem, but we will not attempt that in the present paper.

Conceptually, however, it is fairly easy to understand that a zero mode of $\psi$ is only possible if and only if $\tilde S\equiv\frac{S_x}{\mu}<1$. Plugging the solution (\ref{eqn:nbkg}) into the eq. (\ref{schrpot}) the potential takes the form:
\beq
 V_{\rm eff}(r)=-\mu^2 \Big( \big(1-\frac{1}{r}\big)^2-\tilde S^2\frac{f}{r^2}\Big)-f^2 \Big(-\frac{(d-1)(d-3)}{4r^2}-\frac{(d-1)f'}{2rf}-{m^2 \over f}\Big)\,,
 \label{potax}
\eeq
which, for simplicity, can be rewritten as
\beq
V_{\rm eff}(r)=\mu^2{\cal V}+V^{0}_{eff}\;,\quad {\rm where}
\quad {\cal V}=-\Big( \big(1-\frac{1}{r}\big)^2-\tilde S^2\frac{f}{r^2}\Big)\,,
\label{potsplit}
\eeq
and $V^{0}_{eff}$ is the effective potential for the scalar in the absence of gauge fields. Considering the fact that $\cal V$ is a simple polynomial in $1/r$, it may be shown that for $\tilde S\geq 1$, ${\cal V}$ is non-negative. Hence $V_{\rm eff} \geq V^{0}_{eff}$ and we conclude that the lowest eigenvalue of $\tpsi$ in the potential $V_{\rm eff}$ is strictly greater than the lowest eigenvalue of $\tpsi$ in the potential $V^0_{\rm eff}$. In particular, $V_{\rm eff}$ cannot have a bound state if $V^{0}_{\rm eff}$ does not already have one. Then for $\tilde S \geq 1$ there cannot be an instability towards the condensation of $\psi$.
In the case $\tilde S<1$, it is easy to see that the lowest value of ${\cal V}$ is $\tilde S^2-1$, occurring at the boundary $r \rightarrow \infty$. Moreover, for sufficiently large values of $\mu$, $V_{\rm eff}$ becomes negative enough in a region of finite support near the boundary. Hence it is natural to expect a condensation of $\psi$. These observations are indeed supported by our numerics.

Unfortunately, the EOM of $\psi$ is not exactly solvable in general. Numerically one may find out the trajectory in the $(\tilde S,\frac{1}{\mu})$ plane where the zero mode of $\psi$ occurs. However, even with $S_x=0$ the EOM of $\psi$ is not solvable in general. Although an exact solution near the phase transition for $m^2=-4$ in $d=4$ has been presented in \cite{1003.3278}, till date there is no known exact solution with non-zero $S_x$. Consequently there is not much analytical understanding of the nature of the phase transition at finite $S_x$. Here we will show that taking a certain double scaling limit enables us to exactly solve the EOM for $\psi$. We will thus prove some interesting analytic results about the phase transition using this scaling limit. But before doing that we present the detailed numerical solution of eq. (\ref{seom}) for various specific cases and discuss the corresponding phase structure.

\subsection{$d=3$}

\begin{figure}
\centering
\hspace{-1cm}\begin{minipage}[b]{.5\textwidth}
\centering
\includegraphics[width=7cm]{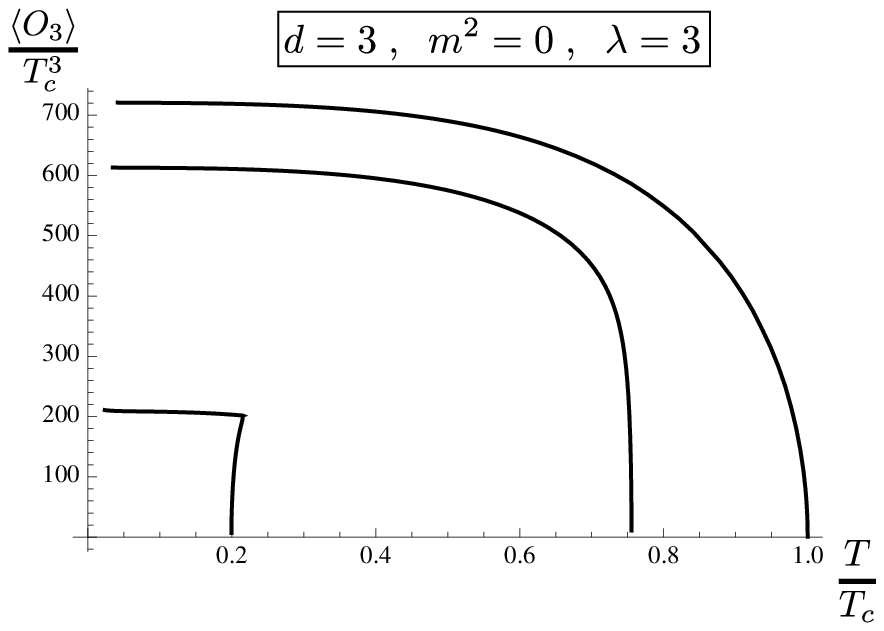}
\end{minipage}%
\begin{minipage}[b]{.55\textwidth}
\centering
\includegraphics[width=7cm]{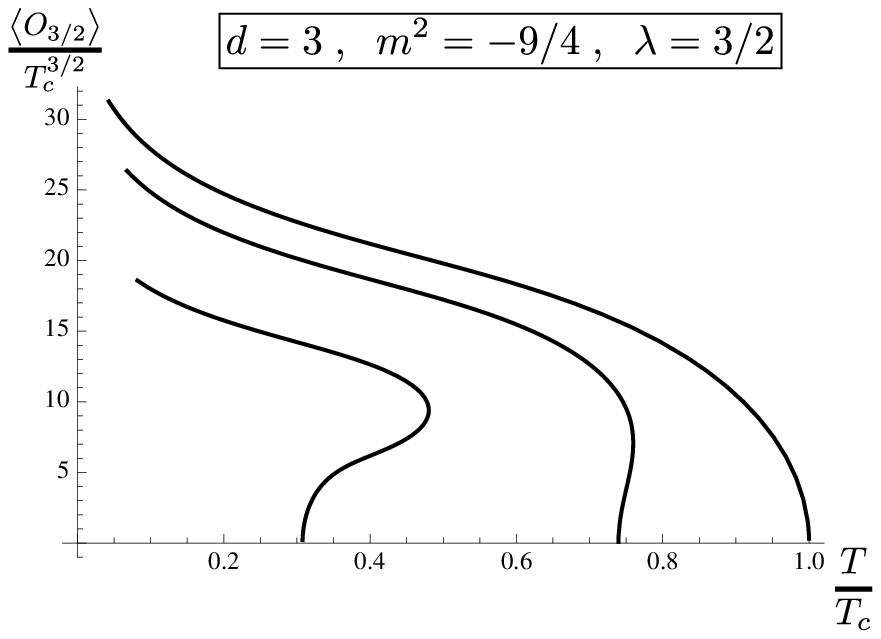}
\end{minipage}\\[-10pt]
\hspace{-1.4cm}\begin{minipage}[t]{.4\textwidth}
\caption{
$\tilde S=0.0, 0.32, 0.75$ increasing to the left.
}\label{d3m0}
\end{minipage}%
\hspace{1.8cm} \begin{minipage}[t]{.4\textwidth}
\caption{
$\tilde S=0.0, 0.36, 0.64$ increasing to the left.
} \label{d3m-9by4}
\end{minipage}%
\end{figure}

\begin{figure}
\centering
\hspace{-1cm}\begin{minipage}[b]{.5\textwidth}
\centering
\includegraphics[width=7cm]{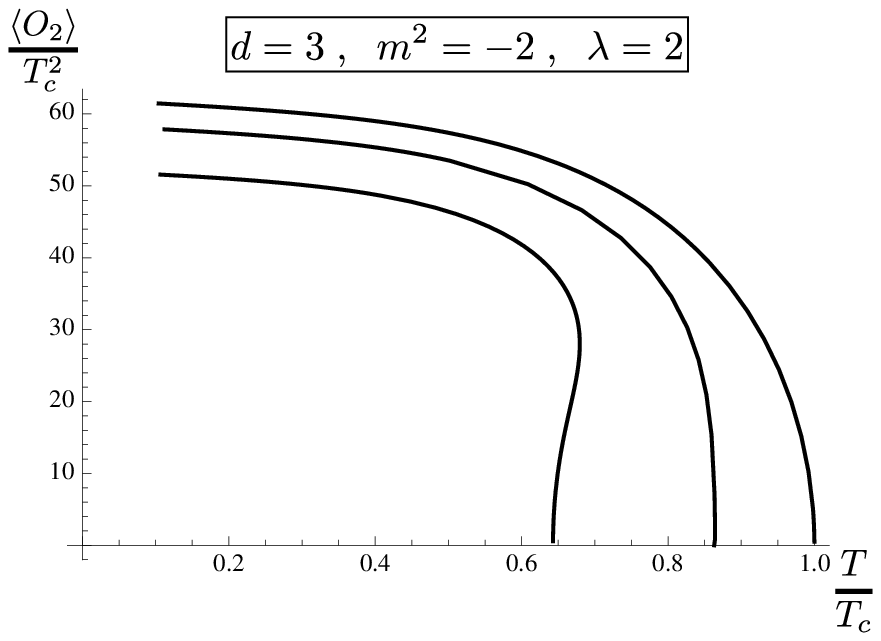}
\end{minipage}%
\begin{minipage}[b]{.55\textwidth}
\centering
\includegraphics[width=7cm]{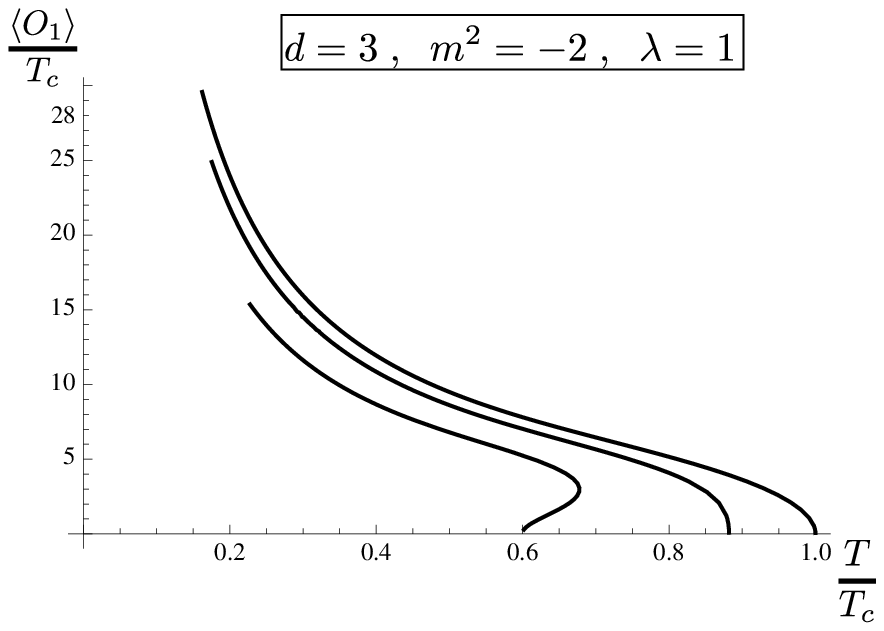}
\end{minipage}\\[-10pt]
\hspace{-1.4cm}\begin{minipage}[t]{.4\textwidth}
\caption{
$\tilde S=0.0, 0.24, 0.40$ increasing to the left.
}\label{d3m-2l2}
\end{minipage}%
\hspace{1.8cm} \begin{minipage}[t]{.4\textwidth}
\caption{
$\tilde S=0.0, 0.30, 0.52$ increasing to the left.
}
\label{d3m-2l1}
\end{minipage}%
\end{figure}

\begin{figure}
\begin{center}\hspace{-0.6in}
\includegraphics[width=0.9\textwidth,
]{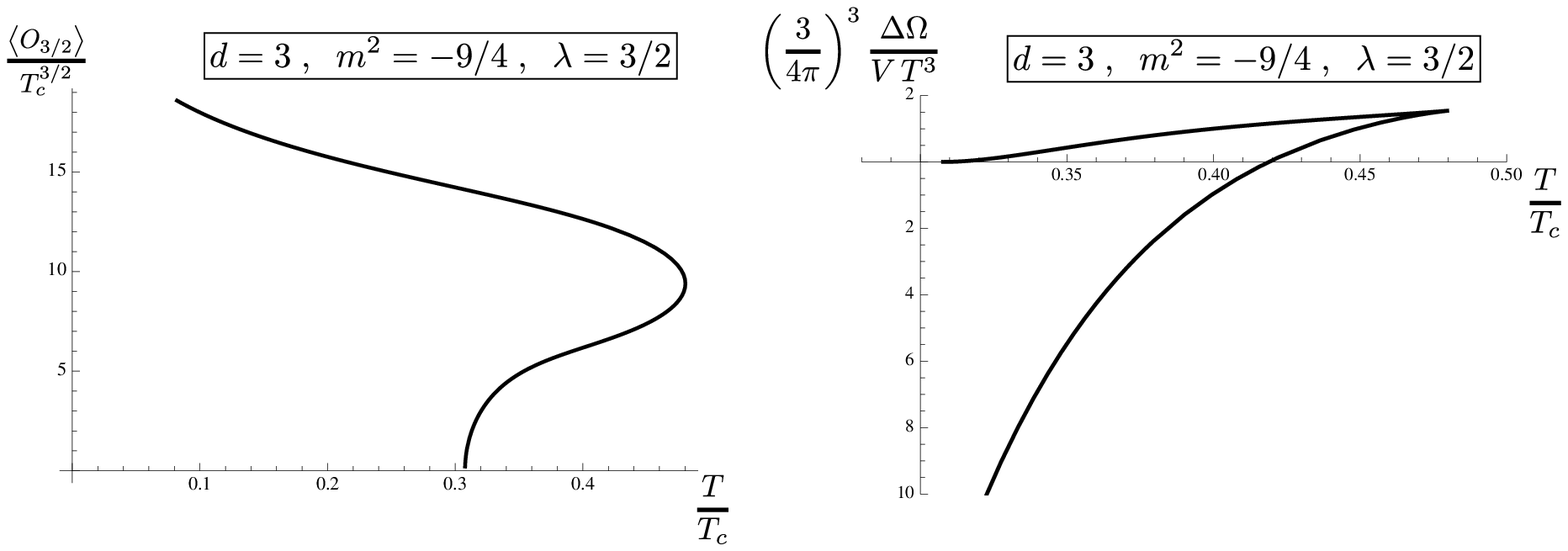}
\caption{A typical first order phase transition: $\tilde S=0.64$ for $d=3,  m^2= -9/4$. On the left we plot the condensate versus the temperature and on the right we show the corresponding free energy
zooming in on the region close to the phase transition.
}
\label{d3m-9by4freeE}
\end{center}
\end{figure}

The phase transitions in the $d=3$ case, for all masses that we have investigated, follow the pattern that was already observed in the conformally coupled $m^2=-2$ case \cite{Basu, Herzog}. At high temperature the only possible phase is the normal phase. At small $\tilde S$ the superconducting phase transition is just like the $S_x=0$ case of \cite{HHH1}. The normal phase becomes locally unstable at $T=\tilde T_c$ (not to be confounded with $T_c$ which all along this paper stands for the critical temperature at $S_x=0$) and a new superconducting phase (I) is created for $T<\tilde T_c$. The superconducting phase has less free energy than the normal phase and dominates over it. This fact may be analytically argued for $S_x=0$. The associated phase transition is of the second order.

$\tilde T_c$ decreases with increasing superfluid velocity and above some critical value of $\tilde S$ the order of the phase transition changes from second to first. This is reflected in the free energy plot (see Figure \ref{d3m-9by4freeE})\footnote{In this plot and the ones of figures \ref{d4m0freeE} and \ref{cow-phase}
we have subtracted the free energy of the normal phase.} as the transition from a smooth curve  to a swallowtail cusp\footnote{A discussion of the free energy is presented in the section \ref{sec:free}.}. As we lower the temperature, before $\tilde T_c$ is reached \footnote{In the cases with many ``critical" temperatures, we use $\tilde T_c$ to denote the phase where the normal phase becomes locally unstable.}, two superconducting phases (apart from the normal phase) become possible at a temperature $T_N$. Unlike the previous case both these phases have a non-zero $\psi$ below $T_N$. The phase with larger value of $\psi$ (phase I) has less free energy than the phase with lower value of $\psi$ (phase II).  The free energy of both superconducting phases (I and II) decreases as $T$ is lowered and
at some temperature, say $T=T_1$, the free energy of phase I becomes lower than the one of the normal phase. Consequently phase I becomes the dominant phase in the system and continues to be so for $T<T_1$. The associated phase transition is of first order.  At a smaller temperature $T=\tilde T_c$ (note that $\tilde T_c<T_1$) the phase II merges with the normal phase.

In a later section, we show by a semi-analytic scaling limit approach that the high-velocity first order transition is present for masses as high as $m^2=20$. We have done detailed direct numerical simulation for a handful of masses ($m^2 \le 4$), some of which we present here in the condensate curves and the free energy plots of figures  \ref{d3m0}, \ref{d3m-9by4}, \ref{d3m-2l2}, \ref{d3m-2l1} and \ref{d3m-9by4freeE}. The phase structure is quite robust.

\subsection{$d=4$}
\label{subsecd4}

\begin{figure}
\begin{center}\hspace{-0.6in}
\includegraphics[width=0.9\textwidth,
]{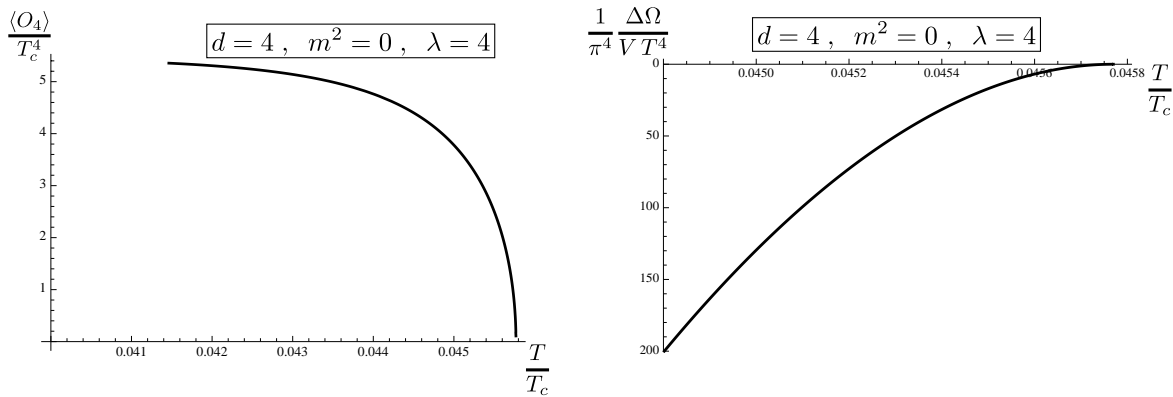}
\caption{A typical second order phase transition: $\tilde S=0.97$ for $d=4,  m^2= 0$.
On the left panel we plot the condensate versus the temperature while on the right panel we present the corresponding free energy, zooming in on the region close to the phase transition.
}
\label{d4m0freeE}
\end{center}
\end{figure}

\begin{figure}
\centering
\hspace{-1cm}\begin{minipage}[b]{.5\textwidth}
\centering
\includegraphics[width=7cm]{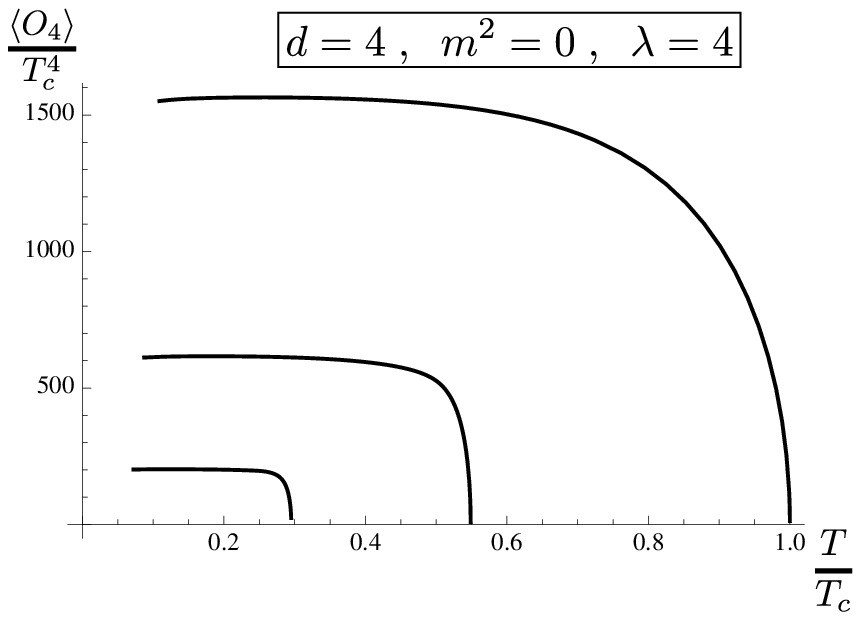}
\end{minipage}%
\begin{minipage}[b]{.55\textwidth}
\centering
\includegraphics[width=7cm]{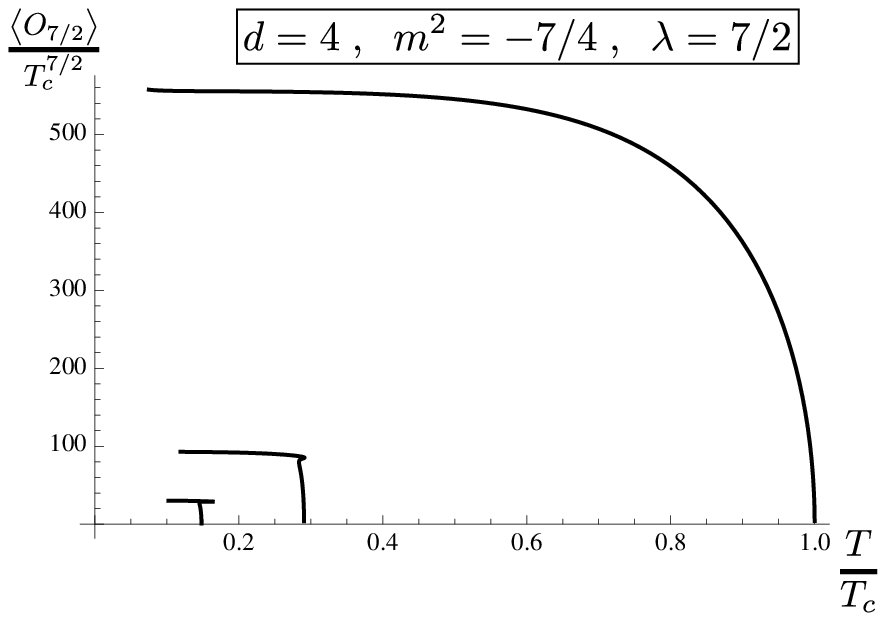}
\end{minipage}\\[-10pt]
\hspace{-1.4cm}\begin{minipage}[t]{.4\textwidth}
\caption{
$\tilde S=0.0, 0.61, 0.8$ increasing to the left.
}\label{d4m0}
\end{minipage}%
\hspace{1.8cm} \begin{minipage}[t]{.4\textwidth}
\caption{
$\tilde S=0.0, 0.8, 0.9$ increasing to the left.
} \label{d4m-7by4}
\end{minipage}%
\end{figure}

The $d=4$ case is more interesting. Here, we find that as the mass of the scalar increases, qualitatively new features appear. In particular, for high enough mass, the phase transition remains second order for arbitrarily high values of $S_x$, as is evident for example from our $m^2=0$ plots in figures \ref{d4m0} and \ref{d4m0freeE}. For small enough mass the phase structure is similar to that of the $d=3$ case. In the intermediate mass scale the phase diagram shows interesting new features.

\begin{figure}
\begin{center}\hspace{-0.6in}
\includegraphics[
width=0.51\textwidth
]{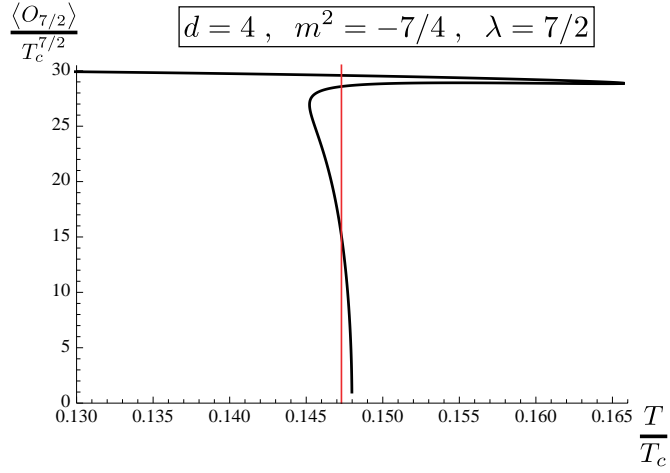}
\caption{Zoom on the $\tilde S=0.9$ case of figure \ref{d4m-7by4}. The vertical (red) line indicates the first order phase transition between two superconducting states characteristic of the Cave of Winds phase structure. 
}
\label{caveofwinds}
\end{center}
\end{figure}

\begin{figure}
\begin{center}
\includegraphics[width=0.9\textwidth,
]{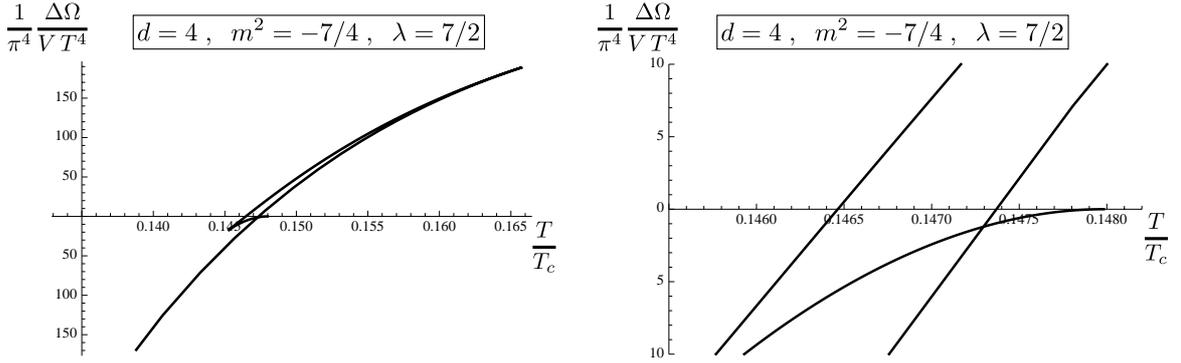}
\caption{Free energy (and its zoom) for $\tilde S=0.9$, $d=4, m^2=-7/4$.
\hspace{1.6in}
}
\label{cow-phase}
\end{center}
\end{figure}

{\bf ``Cave of Winds'':} A numerical scan of all values of the mass is quite challenging, so we have settled for plotting the graphs for a few representative values. A particularly interesting representative case is $m^2=-7/4$ (see figure \ref{d4m-7by4}). Here, for low values of $\tilde S$ the phase transition is always second order. For high enough superfluid velocity, we find that as the temperature is lowered, there is at first a second order phase transition from the normal to a superconducting phase and then a first order phase transition between two superconducting phases (see figures \ref{caveofwinds} and \ref{cow-phase}). We call this the Cave of Winds phase structure because of the shape of the condensate plot.
The basic structure here as we lower the temperature is that initially two new phases (phase I and phase II) appear, both of which are of higher free energy than the normal phase. As we further lower the temperature, a new superconducting phase (phase III)  with lower free energy branches out from the normal phase. At this stage, phase III is the dominant phase. The transition from normal to phase III is second order. At a still lower temperature, there is a first order transition and phase I becomes the dominant phase. 
It is remarkable that the overhang of the condensate curve stretches farther than the location where the transition happens: this is is the reason why the first order transition is to another superconducting phase. A phase transition that was further along the overhang would result in a first order transition directly to the normal phase, and would be in contradiction with the analytical scaling limit results that we report in a later section. Happily, this is not the case. A further comment that is worth making regarding the relation between the analytic scaling limit and direct numerical simulation is as follows. The scaling limit works in a high velocity, low temperature limit and it sees the phase structure close to the normal phase. So it is not surprising that it misses the first order transition between the two superconducting phases that we see via the direct numerical approach. It sees only the second order transition. Another comment is that (as we observed above) the first order phase transition happens long before the tip of the overhang: we have noticed that this is a fairly generic phenomenon that happens for other cases as well, and not just for the Cave of Winds case. In the language of Ginzburg-Landau effective actions, a cave of wind like situation could appear if some higher order term (higher than quartic) in the effective action changes sign: we discuss this in the appendix.

For the even higher mass case of $m^2=0$, the phase transition is always second order for any superfluid velocity. In a later section we investigate various masses in $d=4$ using a high velocity scaling limit, analytically. The results suggest that the critical mass at which the phase transition structure changes is $m^2\approx -2.457$. Below that the phase transition at high velocities is first order. The analytical approach suggests that above that mass the higher temperature phase transition is second order at high velocities, but it is silent about what happens at lower temperatures. Indeed, for the values of mass that we have done detailed numerical simulation,
we find that at high velocities there is always a second order transition. While the analytical scaling limit is rather simple and can be applied to many cases at once, it suffers from the drawback that it cannot be applied to understand phase transitions away from the scaling limit. This means in particular that it only sees phase transitions in which at least one of the phases is the normal phase. Therefore it has nothing to say about (say) the first order transition that we see in the $m^2=-7/4$ case. Of course, in all the cases where we expect overlap between the two approaches, we find matches. There seems to be a lot of structure in the intermediate range of masses close to $m^2 \approx -2.457$ and we hope to come back to a detailed numerical investigation of this range in future work.

\begin{figure}
\centering
\hspace{-1cm}\begin{minipage}[b]{.5\textwidth}
\centering
\includegraphics[width=7cm]{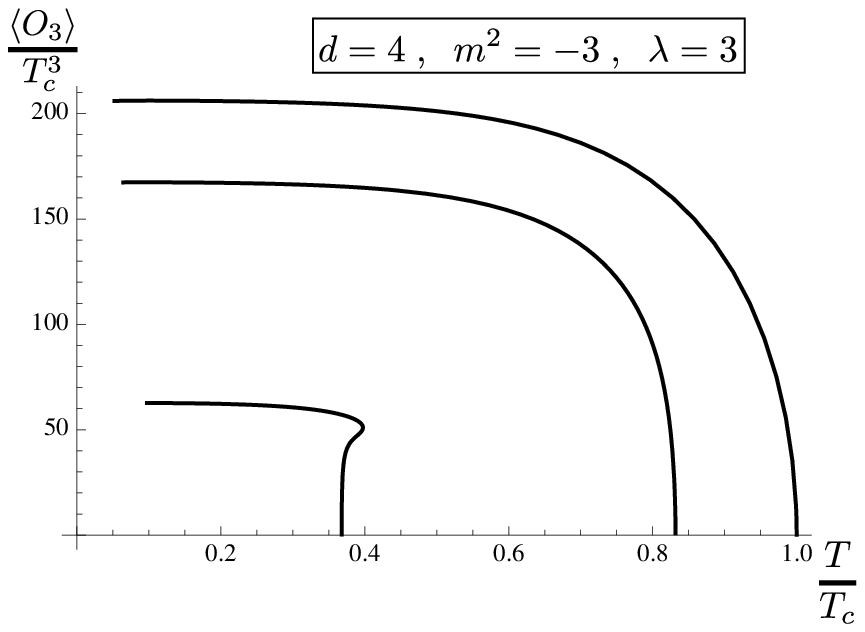}
\end{minipage}%
\begin{minipage}[b]{.55\textwidth}
\centering
\includegraphics[width=7cm]{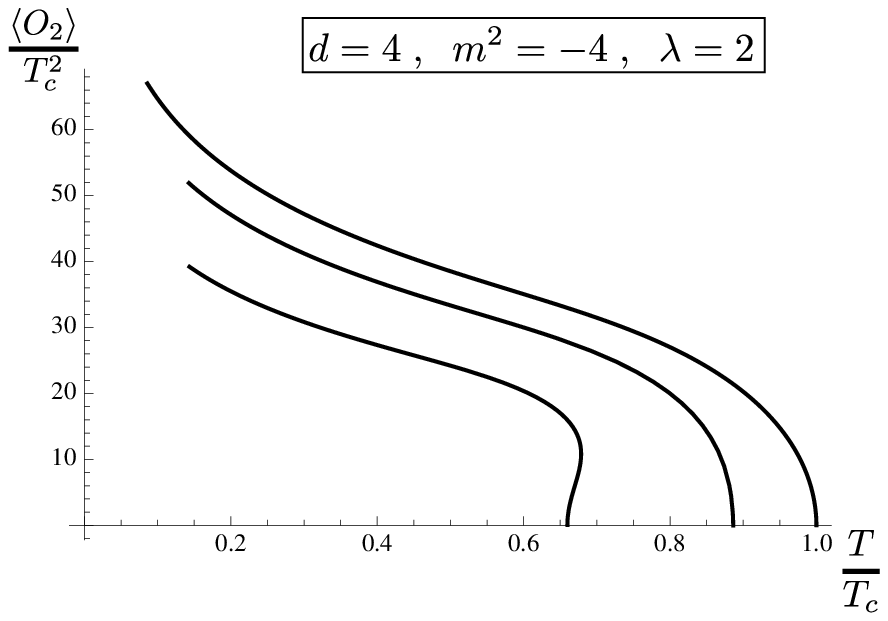}
\end{minipage}\\[-10pt]
\hspace{-1.4cm}\begin{minipage}[t]{.4\textwidth}
\caption{$\tilde S=0.0, 0.36, 0.74$ increasing to the left.}
\label{d4m-3}
\end{minipage}%
\hspace{1.8cm} \begin{minipage}[t]{.4\textwidth}
\caption{$\tilde S=0.0, 0.33, 0.55$ increasing to the left.}
\label{d4m-4}
\end{minipage}%
\end{figure}

\begin{figure}
\centering
\hspace{-1cm}\begin{minipage}[b]{.5\textwidth}
\centering
\includegraphics[width=7cm]{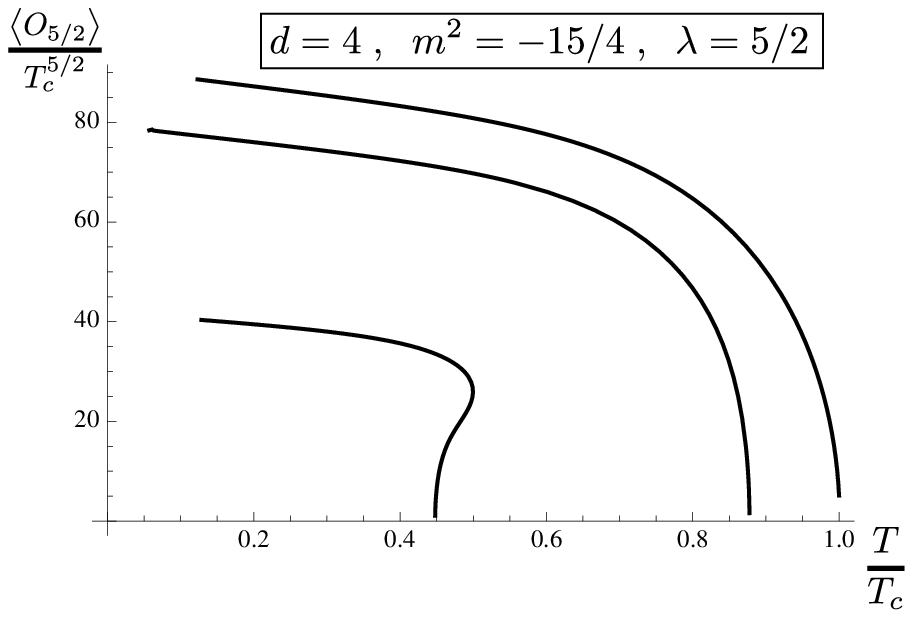}
\end{minipage}%
\begin{minipage}[b]{.55\textwidth}
\centering
\includegraphics[width=7cm]{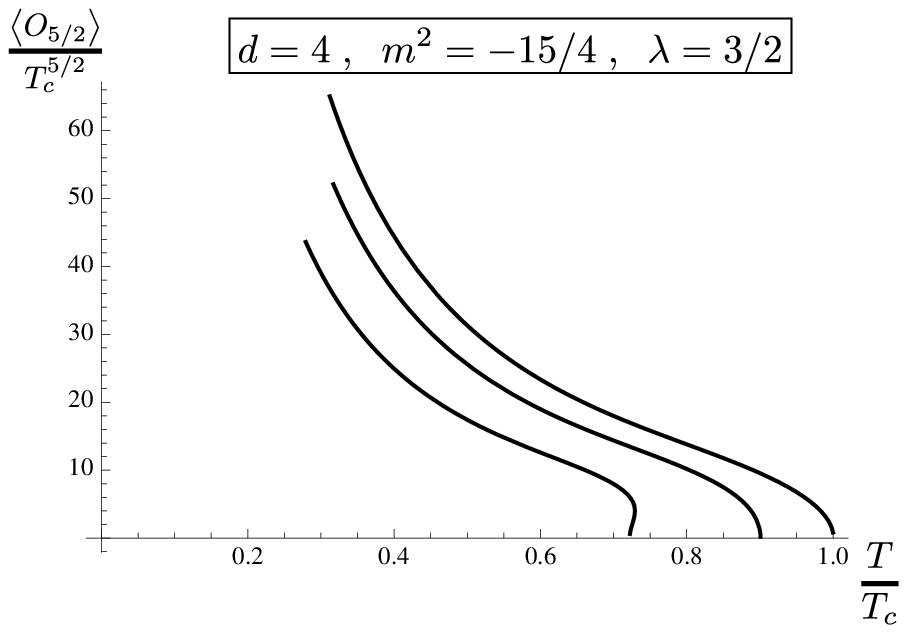}
\end{minipage}\\[-10pt]
\hspace{-1.4cm}\begin{minipage}[t]{.4\textwidth}
\caption{
$\tilde S=0.0, 0.32, 0.68$ increasing to the left.}
\label{d4m-15by4l5by2}
\end{minipage}%
\hspace{1.8cm} \begin{minipage}[t]{.4\textwidth}
\caption{
$\tilde S=0.0, 0.35, 0.56$ increasing to the left.}
\label{d4m-15by4l3by2}
\end{minipage}%
\end{figure}

Finally, in figures \ref{d4m-3}, \ref{d4m-4}, \ref{d4m-15by4l5by2} and \ref{d4m-15by4l3by2} we show the results of our numerical analysis for the cases $m^2=-3$, $-4$ and $-15/4$ in $d=4$. As mentioned above, for these low values of the mass we find the same phase structure as in the $d=3$ case.

\subsection{Free Energy}\label{sec:free}

To understand the nature of the phase transitions, we now describe the computation of the free energy. We will plot the free energy of the normal phase and the superfluid phase for various values of the superfluid velocity $S_x$ to corroborate our claim that the phase transition changes from second order to first order as we increase $S_x$.

The free energy of a solution is defined essentially as the on-shell action after the appropriate addition of boundary (counter) terms: $\Omega=-T S_{\rm os}$. Substituting our ansatz in terms of $A_t(r), A_x(r), \psi(r)$ and the background (\ref{BHmetric}) in the action of the Einstein-Maxwell-scalar theory given by eq. (\ref{qaction}) we get
\beq
S_{\rm bare}=-\int d^{d+1}x\ r^{d-1}\left(-\frac{A_t'^2}{2}+\frac{f}{2r^2}A_x'^2+m^2 \psi^2-\frac{1}{f}A_t^2\psi^2+\frac{1}{r^2}A_x^2 \psi^2+f \psi'^2\right)\,,
\label{Sans}
\eeq
and after using the EoM's (\ref{seom}) we arrive at
\beq
S_{\rm os}=\int d^{d}x\left( \frac{r^{d-1}}{2}  A_t A_t'-\frac{r^{d-3}}{{2}} f A_x A_x'-f r^{d-1}\psi \psi ' \right)\Big|_{r=\infty}+\int d^{d+1} x \left(r^{d-3}\psi^2A_x^2-\frac{r^{d-1}}{f}\psi^2A_t^2\right).\hspace{0.1in}
\label{Sos}
\eeq
Notice that we have both boundary and bulk terms contributing to the on-shell action and the bulk contributions are always finite in the cases we consider here. From the falloffs (\ref{falloff1}, \ref{falloff2}), we find that the only divergences possible are from the scalar terms in the boundary action. The scalar terms are schematically of the form
\bea
S^{\psi}_{\rm boundary} \sim \lambda_- \psi_-^2 \ r^{\sqrt{d^2+4 m^2}}+(\lambda_-+\lambda_+) \psi_+\psi_-+\lambda_+\psi_+^2 \ r^{-\sqrt{d^2+4 m^2}}+...
\eea
where $r$ is the radius at which we cut-off the geometry\footnote{This expression is written for the case when the $\lambda$'s are not equal.  But the result for the final expression of the finite part holds even in that case.}. The first piece is divergent and is killed off by the counter-term \cite{Daniel, Ross}. The final piece is zero when we take the cut-off to infinity. The finite piece is the middle piece and its coefficient gets corrected due to the finite part of the counter-term. Even the remaining finite part will not contribute to the free energy in our case because one of the two among $\psi_{\pm}$ is always zero for us. The specific counter-term is chosen on the basis of the boundary conditions and ensemble that we are working with, see \cite{Herzog} and section 3.1.1 of \cite{Daniel} for a discussion. In our case here, we are working with superfluid velocity fixed as opposed to the current fixed.

The upshot of the discussion above is that the scalar pieces do not contribute in our computation of the free energy.
Then, the final expression of the free energy that we will use in our numerical analysis is
\bea
\frac{\Omega}{T^d V}=\left({4\pi\over d}\right)^d\left[-\frac{d-2}{2}\,(A_t^{(0)}\,A_t^{(d-2)}-A_x^{(0)}\,A_x^{(d-2)})-\int dr \left( r^{d-3}\psi^2A_x^2-\frac{r^{d-1}}{f}\psi^2A_t^2 \right)\right]\,,
\label{freeEnum}
\eea
where $(A_t^{(i)}, A_x^{(i)})$ correspond to the falloffs of the dimensionless rescaled gauge fields
\beq
A_t=A_t^{(0)}-\frac{A_t^{(d-2)}}{r^{d-2}}+...\;,\qquad
A_x=A_x^{(0)}-\frac{A_x^{(d-2)}}{r^{d-2}}+...
\label{dimlessfalloffs}
\eeq
and are related to the dimensionful falloffs of eqs. (\ref{falloff1}, \ref{falloff2}) through the rescaling (see eq. (\ref{r0scaling}))
\beq
\mu=r_0\,A_t^{(0)}\;,\qquad \rho=r_0^{d-1}\,A_t^{(d-2)}\;,\qquad
S_x=r_0\,A_x^{(0)}\;,\qquad  J_x=r_0^{d-1}\,A_x^{(d-2)}\;.
\label{rescrel}
\eeq
The $T^d$ and the overall factor $(4\pi/ d)^d$ in (\ref{freeEnum}) arise precisely because we are working with dimensionless variables to perform the numerics.
We have also divided by the volume of the boundary. As we have mentioned in footnote 8 in the free energy graphics of figures \ref{d3m-9by4freeE}, {\ref{d4m0freeE} and \ref{cow-phase} we plot $\Delta\Omega$: the free energy of the phase with condensate minus the free energy of the phase without condensate, {\em i.e.} the normal phase corresponding to the analytic solution of eq. (\ref{eqn:nbkg}).


\subsection{Nature of the phase transition}

Let us try to have some analytic understanding of the nature of phase transition. From eq. (\ref{Sans}) by integration by parts and using the scalar EoM (\ref{psieom}) we get
\bea
 \frac{\Omega}{T^d} &=&
 \int d^{d}x \left( -f r^{d-1}\psi \psi ' \right)\Big |_{r=\infty}+\int d^{d+1}x\ r^{d-1}\left(-\frac{A_t'^2}{2}+\frac{f}{2r^2}A_x'^2 \right) \nonumber \\
 &=&\int d^{d+1}x\ r^{d-1}\left(-\frac{A_t'^2}{2}+\frac{f}{2r^2}A_x'^2 \right),
\label{eqn:newform}
\eea
where in the last line we have used the fact that the boundary value of $\psi$ is kept to zero in our case. Writing the action in this simple form gives us information about the relative free energy of the different phases.

The free energy for the normal phase is given by $\frac{\Omega_{normal}}{T^d V}=-(d-2)^2 \mu^2/2$, as one can easily check by plugging the hairless solution (\ref{eqn:nbkg}) into eq. (\ref{eqn:newform}). As discussed before, as we increase the value of $\mu$, $\psi$ condenses and there is a new superconducting phase. The outstanding questions are: (1) Which phase dominates? (2) What is the nature of the associated phase transition? For $S_x=0$ this question may be answered without getting into too much detail. With a chemical potential fixed to $\mu$, the solution of $A_t$ in the new phase can be written as
\bea
A_t=\mu (1-\frac{1}{r^{d-2}}) + \delta A_t\,,
\eea
where $\delta A_t \rightarrow 0$ at $r \rightarrow 1$ and $r\rightarrow \infty$. Then, from eq. (\ref{eqn:newform}) we get,
\bea
\frac{\Omega_{new}}{T^d V} &=& -\frac{\mu^2(d-2)^2}{2}+(d-2)\int dr\, \partial_r(\delta A_t)-\int dr\; r^{d-1} \frac{(\delta A_t)'^2}{2} = \nonumber\\
&=& -\frac{\mu^2(d-2)^2}{2}+(d-2) \delta A_t\Big |_{0}^{\infty}-\int dr \;r^{d-1} \frac{(\delta A_t)'^2}{2} =\nonumber \\
&=& -\frac{\mu^2(d-2)^2}{2}-\int dr \; r^{d-1} \frac{(\delta A_t)'^2}{2}\,.
\eea
Hence
\beq
\delta \Omega= \Omega_{new}-\Omega_{normal} = -(T^d V) \int dr \; r^{d-1} \frac{(\delta A_t'^2)}{2} < 0\,.
\label{eqn:at}
\eeq
Therefore if a phase with non-trivial $\psi$ exists it will always have a lower free energy than the normal phase.

A little extension of the above argument determines the nature of the phase transition {\em locally}. As we increase $\mu$ for fixed $\tilde S$, a zero mode of $\psi= \psi_0$ forms for $\mu=\mu_c$. Following the philosophy of \cite{1003.3278,Herzog:2009ci}, we may consider turning on a small amount of zero mode and at the next order look at the backreaction of this mode on $A_t$ and $A_x$. This enables us to calculate the free energy of the new phase with non-trivial $A_t$ and $A_x$. Let us start by choosing
$\psi=\epsilon\,\psi_0+o(\epsilon^2)$, where $\epsilon$ is a small constant. This perturbation takes the system away from the normal phase to the superconducting phase. We can plug in this value of $\psi$ and solve for $\delta A_t$ and $\delta A_x$ with appropriate boundary conditions. It is clear from the equations that the variation of $A_t$ and $A_x$ will be proportional to $\epsilon^2$, {\em i.e.}
\begin{equation}
\begin{split}
A_t &= A^{(0)}_t+  \epsilon^2 \delta A_t +o(\epsilon^4),\quad A^{(0)}_t =\mu\Big(1-\frac{1}{r^{d-2}}\Big) \\
A_x &= S_x + \epsilon^2 \delta A_x +o(\epsilon^4)
\end{split}
\end{equation}
where
\begin{align}
\delta A_t &=\mu \int \frac{dr_1}{r_1^{d-1}} \int dr\, r^{d-3} \frac{2\psi^2 r^2}{f(r)}\Big(1-\frac{1}{r^{d-2}}\Big)   \\
\delta A_x&= S_x \int \frac{dr_1}{f(r_1)r_1^{d-3}} \int dr \;2\psi^2 r^{d-3} .
\end{align}

We fix the constants of integrations by the boundary conditions at the horizon, {\em i.e.}  $\delta A_t$ vanishes and $\delta A_x$ is regular. It should be noted that boundary values of $A_t$  and $A_x$ also changes by $o(\epsilon^2)$. With respect to the normal phase, the relative action of the present configuration is given by
\bea
\label{eqn:dom}
\frac{\delta \Omega}{T^d V}&=& \epsilon^4 \frac{1}{2} \Big[-\int dr\, r^{d-1}  \Big((\partial \delta A_t)^2-\frac{f}{2 r^2} (\partial \delta A_x)^2 \Big)+ (d-2) \delta\mu^2 \Big] +o(\epsilon^6)\simeq
\nonumber\\
&\simeq&{\cal S} \epsilon^4 +o(\epsilon^6)\,.
\eea
The sign of  ${\cal S}$ determines the phase transition. If ${\cal S} < 0$, the associated transition is locally second order. If $A_x=0$, this is the only possibility as we saw before. On the contrary, the situation becomes more interesting if ${\cal S} > 0$. This happens for a typical first order transition. Although our small fluctuation analysis does not address the question of global phase structure, a positive $\cal S$ strongly suggests (from a simple Ginzburg -Landau picture) the existence of a first order transition (see appendix). 
Notice that we have some more information we can use: for fixed $S_x$, a very large value of $\mu$ implies that the effect of $A_x$ would be negligible and we expect a similar phase diagram to the $A_x=0$ case. Therefore the dominant phase in the large $\mu$ limit is the superconducting phase with non-trivial $\psi$. But since, as we have just explained,  the superconducting phase never becomes locally dominant over the normal phase, the only way in which it can become dominant is through a first order transition.

\subsection{Double scaling limit and exact solutions}

In this section we will show that in a suitably defined double scaling limit where $\tilde S \rightarrow 1$ and $\mu\rightarrow \infty$, the EOM of $\psi$ simplifies drastically.
In this limit the potential $V_{\rm eff}$ defined in eq. (\ref{schrpot}) is negative only in a very small neighbourhood near the boundary $y=0$ (or $r=\infty$), so all the interesting condensation mechanisms occur there.
Therefore we now expand eq. (\ref{psischreom}) around $y=0$ and keep only the leading terms:
\bea
\frac{d^2}{dy^2}\, \tilde \psi- \frac{m^2+\frac{(d^2-1)}{4}}{y^2}\,\tpsi + \mu^2\left((1-2 y^{d-2})-\tilde S^2\right)\tilde \psi + (m^2+\frac{(d-1)}{2}) y^{d-2} \tilde \psi= 0\,,
\label{eqn:lor}
\eea
where we have taken into account that at leading order in $y$, $f(y) \sim {1 \over y^2}$, $A_t=\mu (1-y^{d-2})$, and $A_x=S_x$.
\footnote{Note that the first correction to $f$ is of $o(y^{d})$ and subleading with respect to the $y$ dependent part of $A_t$ so it can be ignored consistently. On the other hand, the $y$ dependent part of $A_t$  is necessary to generate a non-trivial potential for $\psi$ and survives our double scaling limit.}
Let us now define the following double scaling limit:
\bea
&&\hspace{0.6in}\tilde S \longrightarrow 1\;, \qquad \mu\longrightarrow \infty\;, \nonumber \\
&& \tilde \mu\equiv\mu^{2-2 \alpha} (1-\tilde S^2)  \quad \! \text{kept fixed and }\;\; \alpha=\frac{2}{d}\,.
\label{eqn:scaling}
\eea
In this limit equation (\ref{eqn:lor}) reduces to
\begin{align}
\frac{d^2}{dz^2} \tilde \psi - \frac{m^2+\frac{(d^2-1)}{4}}{z^2} \tpsi+ \Big(\tilde \mu-2 z^{d-2}\Big)\tilde \psi = 0,
\label{eqn:lor2}
\end{align}
where we have introduced $z=y \mu^{\alpha}$.
This equation happens to be exactly solvable in $d=4$ for general mass, while the conformal case ($m^2=-\frac{(d^2-1)}{4}$) is also solvable in $d=3$. We shall first focus on the $d=4$ case.

\subsubsection{$d=4$}

In $d=4$, we can solve eq. (\ref{eqn:lor2}) exactly. Generically this equation may be solved in terms of a product of Laguerre polynomials and Gaussians. However we will not present the full solution here. Our interest is confined to finding a node-less, normalizable solution. Such a solution is given by
\beq
\tpsi(z)= z^{\tlambda} \exp(-\frac{z^2}{\sqrt 2})\,,
\label{eqn:ssol}
\eeq
with
\beq
\tlambda(\tlambda -1)=m^2+\frac{d^2-1}{4}.
\label{tdeltadef}
\eeq
In addition, we must require the following relation between $\tlambda$ and $\tilde \mu$:
\begin{align}
\tilde \mu_c&=\sqrt{2} (2\tlambda+1).
\end{align}
It might seem surprising that we are demanding that there be a constraint on the parameters of the differential equation that we started with. The appropriate comparison here is with the energy $E$ in the standard Schr\"odinger equation of quantum mechanics, which can get quantized due to boundary conditions. It should be noted that eq. (\ref{eqn:lor2}) does not always have normalizable solutions. These only occur for the specific values of $\tilde \mu(=\tilde \mu_c)$ given by the above relation. These are then the values of $\tilde \mu$ for which we may find zero modes of $\psi$.

As follows from eq. (\ref{tdeltadef}) $\tlambda$ has two solutions for each value of $m^2$:
\beq
\tlambda=\tlambda_{\pm}=\frac{1}{2}(1 \pm \sqrt{4m^2+d^2}).
\eeq
It should be kept in mind that as follows from eq. (\ref{psicov}) $\psi\sim y^{\frac{d-1}{2}}\tpsi$ near the boundary ($y\sim0$). Consequently, using eq. (\ref{eqn:ssol}) $\psi \sim z^{\lambda} \sim z^{\tlambda+\frac{d-1}{2}}$ (at $z\sim0$). Notice that $\lambda=\tlambda+(d-1)/2$ is the dimension of the operator dual to $\psi$ and has the two possible values written in eq. (\ref{falloff1}).
The unitarity bound gives the constraint $\tlambda>-\frac{1}{2}$ \cite{KW2}. Thus $\tlambda_{+}$ is always a valid choice for $\tlambda$ and for $-\frac{d^2}{4} < m^2 < -\frac{d^2}{4}+1$, $\tlambda_{-}$ is also a valid choice. In the table \ref{tab:action}, we have summarized the properties of the system for a few specific values of $\tlambda$.

It is straightforward to address the eigenvalue problem associated with eq. (\ref{eqn:lor2}).
For $\tilde \mu > \tilde \mu_c$, the zero mode discussed here becomes a negative energy bound state and $\psi$ tends to condense. No such instability exists for $\tilde \mu < \tilde \mu_c$.

Besides the ground state there exist excited solutions with nodes in the $z$ direction. These solutions are given in terms of Laguerre polynomials. We will not explore them in full detail and only write down the solution for $m^2=-\frac{15}{4}$:
\begin{align}
\tpsi(z)={\text H}_n(2^{\frac{1}{4}}z) \exp(-\frac{z^2}{\sqrt 2}),\quad \tilde \mu=(2n+1)\sqrt{2},\quad  n \in \mathbb{Z^{*}}\,.
\end{align}
Furthermore, $\tlambda=1$ implies an odd $n$ and $\tlambda=0$ implies an even $n$.

\subsubsection{$d=3$}

Here we will look at the conformal case {\em i.e.}, $m^2=-2$. Normalizability at $z=\infty$ fixes
\begin{align}
\tpsi = \text{AiryAi}\Big(\frac{-\tilde \mu +2 z}{2^{\frac{2}{3}}}\Big)\,,
\end{align}
up to a multiplicative constant.
There may be two possible boundary conditions near the boundary $z=0$. We may choose either $\tpsi(0)=0$ or $\tpsi'(0)=0$. They correspond to the condensation of the two possible dual operators. Choosing the appropriate boundary condition fixes the value of $\mu$ for which a zero mode of $\tpsi$ exists.

The boundary condition $\tpsi(0)=0$ implies
\beq
\tilde \mu_c= 2^{\frac{2}{3}} z_0\,,
\label{eqn:airzero}
\eeq
where $z_0$ is some zero of the $\text{AiryAi}(z)$. It should be noted that all zeros of $\text{AiryAi}(z)$ lie in the $z<0$ region. Choosing the first zero (nearest to $z=0$) gives us a solution which does not have a node. Choosing other zeros give rise to multi-nodal solutions \cite{Gubser:2008zu}. It is believed that these solutions always have higher free energy than the node-less soliton and do not play a significant role in phase transition. Choosing the first zero of $\text{AiryAi}(z)$ we get
\begin{align}
\tpsi(0)=0 &\Rightarrow  \tilde \mu_c \approx 3.71151
\end{align}

On the other hand, the boundary condition $\tpsi'(0)=0$ implies
\beq
\tilde \mu_c= 2^{\frac{2}{3}} z_0\,,
\label{eqn:airzero}
\eeq
where $z_0$ is some zero of the $\text{AiryAiPrime}(z)=\frac{d}{dz}\text{AiryAi}(z)$. Similarly to the previous case, all zeros of $\text{AiryAiPrime}(z)$ lie in the $z<0$ region. The first zero (the one nearest to $z=0$) gives us a solution which does not have a node. Choosing other zeros gives rise to multinodal solutions as before. For a node-less solution
\begin{align}
\tpsi'(0)=0 &\Rightarrow \tilde \mu_c \approx 1.61723
\end{align}

\subsubsection{Free energy in the double scaling limit}

Here we will concentrate mainly on the $d=4$ case. We will follow the general philosophy of section \ref{sec:free} to calculate the free energy in a small fluctuation analysis. Our calculation would be in the double scaling limit (\ref{eqn:scaling}) where $\tilde S\to1$.  In the scaling limit we may simply take $\frac{f(r)}{r^2}=1$ (we are in the region near the boundary) . Changing to the radial variable $y$ and keeping the leading order terms in $\epsilon$ we get from eq. (\ref{eqn:dom}),
\bea
\delta \Omega&=& V\, \epsilon^4 \mu^2 \bigg\{  \int_{\infty}^{0} dy\, y^{d-3} \Big[\int_{\infty}^{y} dx\;\tpsi^2 (1-x^{d-2}) \Big]^2 -\tilde S^2 \int_{\infty}^{0} dy\, y^{d-3} \Big[\int_{\infty}^{y} dx\;\tpsi^2 \Big]^2-
\nonumber\\
&&-\,(d-2)\left[\int_{\infty}^{0}dy\, y^{d-3} \left(\int_{\infty}^{y}  dx\;\tpsi^2 (1-x^{d-2}) \right)\right]^2  \bigg\}\,.
\eea
which in terms of the rescaled variable $z=y \mu^\frac{2}{d}$ becomes
\bea
\delta \Omega&=&2 V\,\epsilon^4 \mu^\frac{4-2d}{d} \bigg\{\tilde \mu_c \int_{\infty}^{0} dz\, z^{d-3} \Big[\int_{\infty}^{z} dx\;\tpsi^2 \Big]^2 -2    \int_{\infty}^{0} dz\; z^{d-3} \Big[ \int_{\infty}^{z} dx\;x^{d-2} \tpsi^2 \int_{\infty}^{z} dy\; \tpsi^2 \Big] - \nonumber\\
&&(d-2) \left[\int_{\infty}^{0}dz\, z^{d-3} \left(\int_{\infty}^{z} dx\;  \tpsi^2 \right)\right]^2 \bigg\}\,.
\label{integral}
\eea
Here we have kept only the leading terms in $\mu$. This expression is exact in the double scaling limit (\ref{eqn:scaling}).

In $d=4$ the above integral may be evaluated analytically using the exact solution (\ref{eqn:ssol}). We get
\bea
\frac{\delta \Omega}{V\,T^3 \epsilon^4 \mu^{-1}} = 2^{-3 \tlambda -\frac{11}{2}} \left[4^{\tlambda } \left(4 \tlambda
   ^2-1\right) \Gamma \left(\tlambda +\frac{1}{2}\right)^2-8 \tlambda\,\Gamma\, (2 \tlambda +1)\right]\,.
\label{eqn:anex}
\eea
The Gamma function can be expressed in terms of integers, radicals and $\pi$'s when $\tlambda$ is an integer or a half integer. In table \ref{tab:action} we summarize a few values of $\tlambda$ and the corresponding values of $\delta \Omega$. We also plot the dependence of $\delta \Omega$ on $\tlambda$ numerically (figure \ref{actfig}). For small values of $\tlambda$, $\delta \Omega$ is positive and it becomes negative for large values of $\tlambda$. It may also be argued using Sterling's approximation that $\delta \Omega$ is a negative quantity in the large $\tlambda$ limit. $\delta \Omega$ changes sign between $\tlambda=\frac{3}{2}$ and $\tlambda=2$, to be precise at $\tlambda \approx 1.742$ (corresponding to $m^2\approx-2.457$).

As we discussed before, a positive value of $\delta \Omega$ indicates that locally the superconducting phase has more free energy than the normal phase and there is no second order transition from the normal to the superconducting phase. However this opens up the possibility of a global phase transition between the normal and superconducting phases. Considering that for $\tilde \mu>\tilde \mu_c$ the normal phase is locally unstable, there possibly exists a global transition before $\tilde \mu$ reaches $\tilde \mu_C$.  On the other hand, if $\delta \Omega$ is negative then locally the superconducting phase has lower free energy than the normal phase and there will be a second order phase transition from the normal to the superconducting phase.

\begin{figure}
\begin{center}
\includegraphics[width=0.55\textwidth]{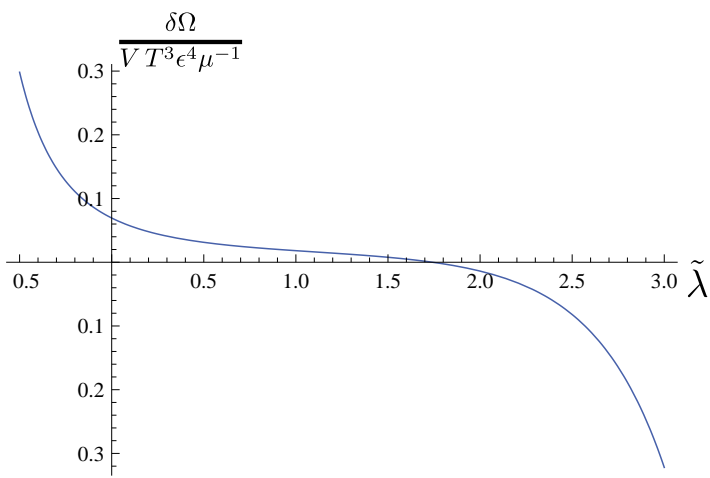}
\caption{Dependence of $\delta \Omega$ on $\tlambda$. A negative value of $\delta \Omega$ implies a second order transition between the superconducting and the normal phase. As we discussed before a positive $\delta \Omega$ gives a strong hint for a first order transition in the system.
}
\label{actfig}
\end{center}
\end{figure}

\begin{table}
\begin{center}
\begin{tabular}{|r|r|r|}
\hline
$\lambda=\tlambda+\frac{3}{2} $ & $m^2$ & $ \frac{\delta \Omega}{V\,T^3 \epsilon^4 \mu^{-1}}$ \\
\hline
$ 1 $ & $-3 $ & $ \frac{1}{8}+\frac{\text{Log}[2]}{4} \approx 0.29827 $ \\
\hline
$ \frac{3}{2} $ & $-\frac{15}{4} $ & $ \frac{\pi }{32 \sqrt{2}} \approx 0.06942 $ \\
\hline
$ 2 $ & $ -4 $ & $ \frac{1}{32} = 0.03125 $ \\
\hline
$ \frac{5}{2} $ & $ -\frac{15}{4} $ & $ \frac{16-3 \pi }{256 \sqrt{2}} \approx 0.0181617 $ \\
\hline
$ 3 $ & $ -3 $ & $ \frac{1}{128} = 0.0078125 $ \\
\hline
$ \frac{7}{2} $ & $ -\frac{7}{4} $ & $ \frac{384-135 \pi }{2048 \sqrt{2}} \approx -0.0138504 $ \\
\hline
$4 $ & $ 0 $ & $ -\frac{21}{256} \approx -0.0820313 $ \\
\hline
$ \frac{9}{2} $ & $ \frac{9}{4} $ & $ \frac{17280-7875 \pi }{16384 \sqrt{2}} \approx -0.321963 $ \\
\hline
$ 5 $ & $ 5 $ & $ -\frac{1251}{1024} \approx -1.22168 $ \\
\hline
$ \frac{11}{2} $ & $ \frac{33}{4} $ & $ -\frac{-1290240+694575 \pi }{131072 \sqrt{2}} \approx -4.81125$ \\
\hline
\end{tabular}
\end{center}
\caption{
In this table we display the values of the free energy for several condensates in $d=4$ in the double scaling limit.
In the first two columns we present the dimension of the operator and the mass of its dual bulk field,
while in the third column we show the corresponding value of the free energy computed using eq. (\ref{eqn:anex}). The free energy changes sign between $\lambda=3$ and $\lambda=\frac{7}{2}$.}
\label{tab:action}
\end{table}

The situation is more complicated in the $d=3$ case. Only a piece of the integral (\ref{integral}) can be computed exactly. In particular, $\delta \mu$ may be calculated exactly, but the other second integrals can only be reduced to first integrals. Unfortunately an analytic evaluation of the resulting first integrals seems unlikely and the final result has to be calculated numerically.

\section{Transverse AC Conductivity}

In this section we discuss the AC conductivity results for two characteristic cases: one where the phase transition changes to first order at high velocity and another where it remains second order all the way. We will only consider the transverse conductivity here: the longitudinal conductivity couples to the other modes and things are more complicated. The idea is to solve the transverse gauge field perturbation $\delta A_y= e^{-i \omega t} A_y(r) $  numerically in the background with the condensate. The equation for it takes the form
\bea
\partial_r ( r^{d-3}f \partial_r A_y )+\Big( \frac{\tilde\omega^2 r^{d-3}}{f}-2 r^{d-3} \psi^2\Big)A_y=0.
\eea
We have defined
\bea
\tilde\omega =\frac{\omega}{r_0}=\frac{4 \pi \omega}{T\,d},
\eea
where the last step uses the relationship between the Hawking temperature of the black brane and its Schwarzschild radius. According to linear response theory, the conductivity is defined by a Kubo formula in terms of retarded current-current correlators, which via the AdS/CFT dictionary can be related to the boundary falloffs of $A_y$
\beq
A_y=A_y^{(0)}+\frac{A_y^{(d-2)}}{r^{d-2}}+...
\label{ayfallof}
\eeq
The final result for the conductivity \cite{Horowitz} in $d=3$ is
\bea
\sigma_y(\omega)=i\frac{A_y^{(1)}}{\omega A_y^{(0)}}.
\eea
and in $d=4$
\bea
\sigma_y(\omega)=\frac{2 A^{(2)}}{i \omega A^{(0)}}+\frac{i\omega}{2}.
\eea

The above result (and the retarded correlator) is obtained via the Schwinger-Keldysh prescription by solving the equation with infalling boundary conditions at the horizon. We do this numerically and plot the results for two representative cases in figures \ref{1}, and \ref{2}. The first corresponds to the case $d=4, m^2=0$ ($\lambda=4$),
which is a case that exhibits second order phase transitions at all velocities. We plot the real part of the
conductivity versus the frequency for different values of the superfluid velocity at a fixed temperature  $T=0.22 T_c$. At large frequencies the behavior is linear while for small enough frequencies ${\rm Re} (\sigma)$ vanishes within our numerical precision, this is the conductivity gap. As expected, when the value of the superfluid velocity is close to its maximum (where the phase transition happens) the gap vanishes. This is consistent with the second order nature of the phase transition, and should be contrasted to the second case (figure \ref{2}) where we plot the conductivity for $d=4, m^2=-15/4$ ($\lambda=5/2$ and $T$ fixed to $0.16 T_c$).
Here, as we increase the velocity the gap decreases, until we reach the maximum velocity where the gap
still has a finite value, consistent with the first order nature of the phase transition at high velocities. Roughly, a first order phase transition means that the condensate has a finite binding energy while second order transition corresponds to the case where the condensate can have arbitrarily low binding energy: so the nature of the phase transitions we observe is consistent with the results for the frequency gap.

\begin{figure}
\centering
\hspace{-1cm}\begin{minipage}[b]{.5\textwidth}
\centering
\includegraphics[width=7cm]{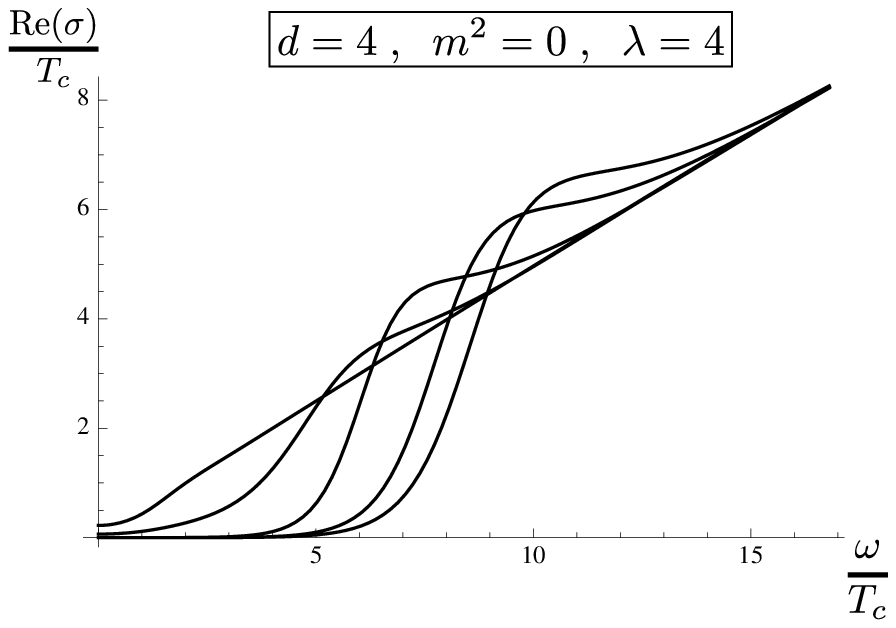}
\end{minipage}%
\begin{minipage}[b]{.55\textwidth}
\centering
\includegraphics[width=7cm]{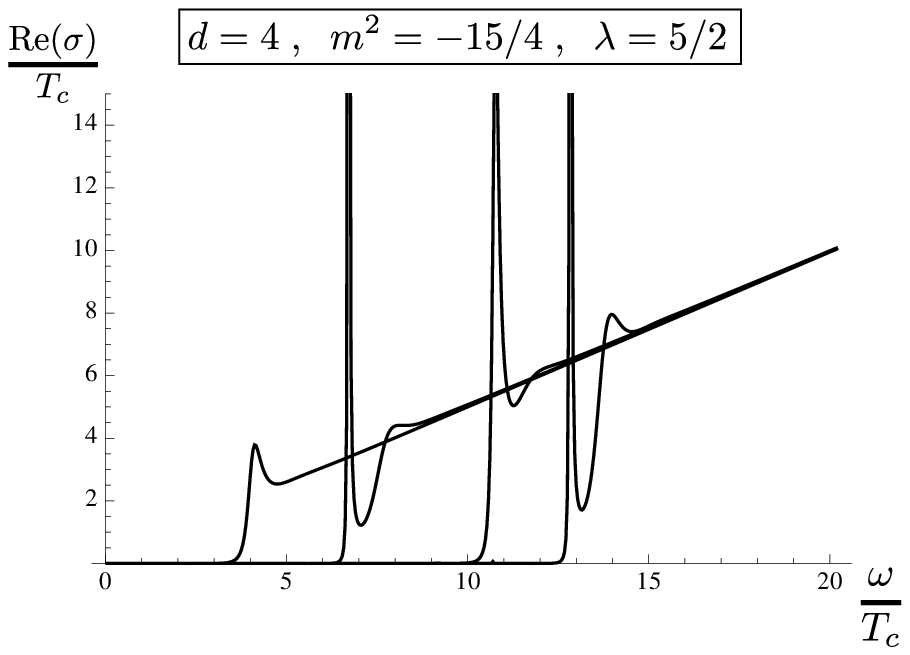}
\end{minipage}\\[-10pt]
\hspace{-1.4cm}\begin{minipage}[t]{.4\textwidth}
\caption{
Plot of ${\rm Re} (\sigma)$ for $d=4$, $m^2=0$ at temperature $T=0.22 T_c$. The velocities are $\tilde S=0.000278, 0.427, 0.708, 0.840, 0.850$, increasing (on the lower part of the figure) to the left.
The highest velocity corresponds to the point where the 2nd order phase transition to the normal phase occurs.
}\label{1}
\end{minipage}%
\hspace{1.8cm} \begin{minipage}[t]{.4\textwidth}
\caption{
Plot of ${\rm Re} (\sigma)$ for $d=4, m^2=-15/4$ at temperature $T=0.16 T_c$. The velocities are $\tilde S=0.000320, 0.505, 0.759, 0.905$, increasing (on the lower part of the figure) to the left. At the highest velocity the system undergoes a 1st order phase transition to the normal phase.} \label{2}
\end{minipage}%
\end{figure}

The study of the AC conductivity will lead us to an interesting observation about the Cave of Winds phase
structure we discussed in section \ref{subsecd4}.
Let us indeed consider the Cave of Winds condensate plots and take two points on either side of the first order phase transition. Note that both these are superconducting states: the one on the upper branch corresponds to what we called phase I before and the one on the lower branch corresponds to phase III. The results for the conductivity are presented in Fig. (\ref{cavegap}) where we clearly see that during the first order phase transition the conductivity gap falls from half of its maximum value (corresponding to $T\to0$) to almost zero. It would be very interesting to see if there are any real material that exhibits such a smoking gun for a ``Cave of Winds" structure: as one lowers the temperature in the superconducting phase, the conductivity gap would jump from  almost zero to half the value at zero temperature.

\begin{figure}
\begin{center}\hspace{-0.6in}
\includegraphics[
height=0.25\textheight
]{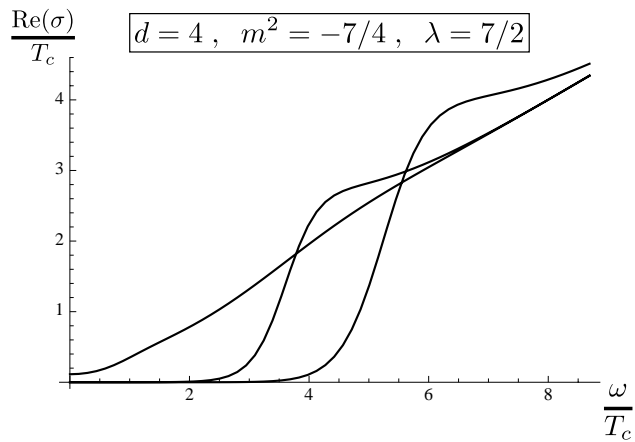}
\caption{ Plot of ${\rm Re} (\sigma)$ for $d=4, m^2=-7/4$ at fixed velocity $\tilde S=0.9$. The different lines correspond to $T$ equal to (from left to right): $0 .147T_c$ (phase III), $0 .147T_c$ (phase I) and  $0.102T_c$ (phase I). \hspace{0.6in}
}
\label{cavegap}
\end{center}
\end{figure}

\section{Summary and Comments}

Our original motivation for the work was that the construction of \cite{Basu, Herzog} was undertaken in the probe approximation where the charge of the scalar $q$ is large and the backreation on the black hole geometry was ignored. This is a somewhat more drastic assumption in the supercurrent case than in the case of the superconducting ground state: the perturbation breaks some of the symmetry of the background, and to take account of a fully backreacted solution we will need to consider a non-static (but stationary) metric. So it is interesting to see whether the results of \cite{Basu, Herzog} where robust.

During the course of our work, \cite{Tisza} appeared, which solved the backreaction problem for the conformally coupled case in $d=3$, and it was found that for small enough values of the charge the phase transition stays second order for all velocities (at least to within numerical control). In the non-backreacted situation that we are considering, we find another kind of deviation from the results of \cite{Basu, Herzog}: when we go to $d=4$, we find that the phase structure can take quite an interesting structure for large enough mass. We saw the emergence of a sequence of phase transitions (of different order) with temperature and also the possibility of a new superconducting phase. Another interesting direction would be to look at zero temperature limit \cite{zero}.

It would be interesting to investigate the phase diagram in more detail in the region of masses near $m^2=-2.457$ in $d=4$, where there seems to be a lot of structure. It might be possible that, beyond what we have already seen, extra features show up in these cases. A plot of our phase diagram where the mass is an extra axis would be interesting because it is not clear how the Cave of Winds phase diagram transitions into the first order structure that we see at high velocities at lower masses. This would require a rather exhaustive numerical scan of the phases.

Another obvious line of development is to consider the backreacted version of some of these cases where one can ask low-temperature questions with more control. This would basically involve the construction of various kinds of fully backreacted hairy black holes in $AdS_5$. In flat 3+1 dimensional spacetime, hairy black holes are ruled out by the famous results of Israel, Carter and Robinson, as well as general arguments by Hawking. In 4+1 dimensions on the other hand, by now it is well known that there is continuous hair \cite{5D}. In AdS as well, it seems likely from our results that the situation in $d=4$ could be richer than what is possible in $d=3$. Yet another natural question is that of the embedding in string theory. Finally, it would also be very interesting to see if there are any real systems in condensed matter which show the kind of Cave of Winds structure that we saw. In particular, the jump in the gap will be an interesting feature of such materials. The fact that we saw this in $d=4$ (and not in $d=3$) suggests that these will probably have to be 3+1 dimensional systems. We hope to come back to some of these problems in the future.

\section*{Acknowledgments}

DA and CK would like to thank M. Bertolini, J. Evslin and T. Prochazka for discussions and J. Sonner and B. Withers for correspondence. PB likes to thank Jianyang He, Anindya Mukherjee and Hsien-Hang Shieh. PB and CK would like to thank the organizers of the Mini-Workshop on String Theory at TIFR, Bombay (Jan 3-8, 2010),  where this project was launched. 
\section{Appendix}
\subsection*{{\bf A.} \ Effective action and phase transition}\label{app:A}
\addcontentsline{toc}{subsection}{{\bf A} \ Effective action and phase transition}
\renewcommand{\theequation}{A.\arabic{equation}}

Here we review some elementary aspects of phase transitions. Let us consider a simple generic effective action
\beq
S= \frac{m^2}{2} \phi^2 + \frac{b}{4} \phi^4 + \frac{c}{6} \phi^6, \quad c>0\,,
\eeq
with an order parameter $\phi$. We choose $c$ to be positive to prevent a runaway situation.

Let us start with a positive $m^2$ and gradually lower it. For a positive $m^2$, the trivial solution $\psi=0$ would be locally stable. As $m^2$ changes sign the trivial solution becomes locally unstable. The mere fact that the trivial solution becomes locally unstable does not automatically imply a second order transition. The order of the transition depends on the sign of $b$. If $b>0$ the phase transition is locally second order. However if $b<0$, then a first order transition happens before $m^2$ changes sign. This feature does not strictly depend on the particular form of our effective action and only relies on the fact that the effective action has no runaway directions and is not unbounded from below.

At the phase transition point there is a zero mode of $\psi$. In our simple case the zero mode is $\phi_0=1$. Near the phase transition point we may turn on a little amount of the zero mode around the trivial solution to perturbatively construct a new solution. Let us assume that $\phi=\epsilon + o(\epsilon^2)$.  Then from the EoM of $\phi$ we find
\beq
m^2=-\epsilon^2 b+o(\epsilon ^4).
\eeq
Hence the on-shell action for the new phase is,
\beq
\delta S=-\frac{b}{2} \epsilon^4+o(\epsilon ^6).
\eeq
Hence the on-shell action of the new phase is related to the sign of $b$ and consequently determines the nature of phase transition. If $b<0$ (or $\delta \Omega>0$), the perturbative phase never becomes a local minimum. Hence when the trivial phase becomes locally unstable, the minimum of free energy lies in some other minimum well separated from the trivial phase. (In the case of a runaway situation such a minimum exists for $\phi \rightarrow \infty$.) This explains why the associated transition would be more abrupt than a second order transition and would generically be first order.

Also to be noted is the parameter range at which the new perturbative phase exists. For $b>0$ the new phase exists for $m^2 < 0$, while for $b<0$ it arises for $m^2 > 0$.

To rephrase the above argument in the language of our problem, we look at the off-shell\footnote{Note that it is the zero mode of the auxiliary Schr\"odinger problem that corresponds to the on-shell case. Here we are considering a non-zero eigenvalue.} effective action of the lowest eigenvalue mode of $\psi$, schematically written as,
\beq
S=m_{\psi}^2 \epsilon^2 + {\cal S} \epsilon^4 + o(\epsilon^5).
\eeq
It may be argued that for small $\mu$, $m_{\psi}^2$ is a positive number. At $\mu=\mu_c$, $m_{\psi}^2$ changes sign. Generically near $\mu=\mu_c$, $m_{\psi}^2=\#(\mu_c-\mu)$. If  ${\cal S} < 0$ a new dominant phase is created for $\mu>\mu_c$. However if ${\cal S }>0$ there is no new phase for $\mu>\mu_c$.

\subsubsection*{The Cave of Winds Structure :}

A Cave-of-Winds-like structure is possible if we include the effect of higher order terms in our effective action. Let us consider an effective action of type
\beq
S= \frac{m^2}{2} \phi^2 + \frac{b}{4} \phi^4 + \frac{c}{6} \phi^6+\frac{d}{8} \phi^8, \quad d>0
\eeq
Now if $c$ changes sign and becomes sufficiently negative, while $b$ remains positive, then there will be a Cave of Wind like structure. Locally near $\phi=0$ the phase transition will always be second order. However because $c$ is negative, there may exist a globally dominant phase which becomes dominant through a first order transition. Whether this phase transition happens before or after the second order transition ({\em i.e.} when $m^2$ crosses zero) is a question of detail.


%
\newpage

\end{document}